\documentclass[]{IEEEtran}

\RequirePackage[normalem]{ulem} 
\RequirePackage{color}\definecolor{RED}{rgb}{1,0,0}\definecolor{BLUE}{rgb}{0,0,1} 

\RequirePackage{color}

\usepackage{tikz}
\usepackage[sumlimits,intlimits]{amsmath}
\usepackage{amsmath,amsfonts,amssymb}%
\usepackage{bm}%
\usepackage{graphicx,graphics}%
\usepackage{cases}%
\usepackage[noadjust]{cite}%
\usepackage{color}%
\usepackage{cite,url}
\usepackage{verbatim}
\usepackage{algorithm}
\usepackage{balance}
\usepackage{multirow}
\usepackage{stfloats}
\usepackage{subfigure}
\usepackage{xspace}
\usepackage{enumerate}

\ifCLASSOPTIONpeerreview
\renewcommand{\baselinestretch}{1.75}
\fi
\begin{document}

\title{\vspace{-20pt}Channel Estimation, Carrier Recovery, and Data Detection in the Presence of Phase Noise in OFDM Relay Systems}
\author{\IEEEauthorblockN{Rui~Wang, Hani Mehrpouyan, \textit{Member, IEEE}, Meixia Tao, \textit{Senior Member, IEEE}, and Yingbo Hua, \textit{Fellow, IEEE}\vspace{-20pt}}

\thanks{R. Wang and M. Tao are with the Department of Electronic Engineering at Shanghai Jiao Tong University, Shanghai,
P. R. China. Emails:\{liouxingrui, mxtao\}@sjtu.edu.cn. H. Mehrpouyan is with the Department of Computer and Electrical Engineering and Computer Science at California State University, Bakersfield, CA, USA. Email: hani.mehr@ieee.org. Y. Hua is with the Department of Electrical Engineering at the University of California, Riverside, CA, USA. Email: hua@ee.ucr.edu.}}

\maketitle

\begin{abstract}
Due to its time-varying nature, oscillator phase noise can significantly degrade the performance of channel estimation, carrier recovery, and data detection blocks in high-speed wireless communication systems. In this paper, we analyze joint channel, \emph{carrier frequency offset (CFO)}, and phase noise estimation plus data detection in \emph{orthogonal frequency division multiplexing (OFDM)} relay systems. To achieve this goal, a detailed transmission framework involving both training and data symbols is presented. In the data transmission phase, a comb-type OFDM symbol consisting of both pilots and data symbols is proposed to track phase noise over an OFDM frame. Next, a novel algorithm that applies the training symbols to jointly estimate the channel responses, CFO, and phase noise based on the maximum a posteriori criterion is proposed. Additionally, a new \emph{hybrid Cram\'{e}r-Rao lower bound} for evaluating the performance of channel estimation and carrier recovery algorithms in OFDM relay networks is derived. Finally, an iterative receiver for joint phase noise estimation and data detection at the destination node is derived. Extensive simulations demonstrate that the application of the proposed estimation and receiver blocks significantly improves the performance of OFDM relay networks in the presence of phase noise.

\end{abstract}

\begin{keywords}
Relay networks, amplify-and-forward (AF), hybrid Cram\'{e}r-Rao lower bound (HCRLB), orthogonal frequency division multiplexing (OFDM), channel estimation, carrier frequency offset, phase noise, receiver design.
\end{keywords}

\vspace{-0pt}
\section{Introduction}

\vspace{-0pt}
\subsection{Motivation and Literature Survey}
Application of relaying has been identified as a suitable approach for combating long-distance channel distortion and small-scale fading in wireless communication systems \cite{Hua_JSAC_12}. Various physical layer techniques, such as distributed space-time block coding \cite{Yin2006}, precoding \cite{RuiWang}, etc., for relay systems have been extensively studied in the past decade. From these works, it can be deduced that to deliver the advantages of relay networks, the network's \emph{channel state information (CSI)} needs to be accurately obtained \cite{Feifei2008, Yindi2012, Lioliou2012, Jun2011, Rong2012}, while the negative impact of impairments such as \emph{carrier frequency offset (CFO)} and \emph{phase noise (PN)} caused by Doppler shifts and oscillator imperfections needs to be mitigated \cite{article_PHYSC_DEAD}.

In single carrier communication systems, CFO and PN are multiplicative and result in a rotation of the signal constellation from symbol to symbol and erroneous data detection
\cite{article_PHASE_N_SISO_EST_BOUNDS, article_SISO_NOELS_IV}.
On the other hand, in the case of \emph{orthogonal frequency division multiplexing (OFDM)} systems, CFO and PN are convolved with the data symbols, resulting in the rotation of the signal constellation and \emph{inter-carrier interference (ICI)}, which can significantly deteriorate the overall performance of an OFDM system \cite{Darryl2006, Darryl2007, article_small_angle_approx_II}. Thus, extensive research has been recently carried out to find carrier recovery schemes that complement traditional approaches, e.g., those based on the \emph{phase-locked loop (PLL)}. More importantly, as demonstrated in \cite{Darryl2006, Mehrpouyan2011}, to accurately obtain the channel, CFO, and PN parameters in communications systems, these parameters need to be jointly estimated. However, the prior art on channel and CFO estimation in relay networks has not taken into consideration the detrimental impact of PN.


%

Due to the presence of multiple hops between source and destination, channel estimation in relay systems is quite different from traditional point-to-point systems. For the amplify-and-forward (AF) relaying strategy, one approach is to only estimate source to destination channels \cite{Feifei2008, Yindi2012}. However, to further enhance cooperative system performance by enabling relay precoding/beamforming or relay resource allocation, the channel response of each hop needs to be separately estimated \cite{Yindi2012, Lioliou2012, Jun2011, Rong2012}. Furthermore, since the channel response from relay to destination affects the destination noise covariance matrix, estimating individual channel responses is generally required for more accurate signal detection at the destination. It is worth noting that the contributions in \cite{Feifei2008, Yindi2012, Lioliou2012, Jun2011, Rong2012} only focus on channel estimation while ignoring the effect of CFO and PN.

Joint estimation of the channel responses and CFO in single carrier relay systems has been considered in \cite{Kyeong2011, Mehrpouyan2011}. In \cite{Kyeong2011}, the Gauss-Hermite integration and approximate Rao-Blackwellization based joint CFO and channel estimators are proposed, while in \cite{Mehrpouyan2011} joint CFO and channel estimation via the MUSIC algorithm is analyzed. However, the works in \cite{Kyeong2011, Mehrpouyan2011} ignore the effect of PN. In fact, although both CFO and PN result in an unknown rotation of signal constellation, PN is a time-varying parameter compared to the CFO and can be more difficult to estimate \cite{article_SISO_NOELS_IV,Darryl2006}. More importantly, the negative impact of CFO and PN may be greater in the case of OFDM systems compared to single carrier systems \cite{Pollet1995, Tomba1998}.

Due to its capability of combating frequency selectivity in the wireless channel, OFDM techniques have been extensively adopted in the latest wireless communication standards, e.g., Long Term Evolution, IEEE 802.11n, Bluetooth, etc. The deteriorating effect of PN on the performance of point-to-point OFDM systems is analyzed in \cite{Pollet1995, Tomba1998}. Undoubtedly, this effect can also be observed in OFDM based corporative relay systems. Hence, conducting accurate channel and CFO estimation in the presence of PN is important for maintaining the quality of service in high-speed OFDM relay networks. Joint estimation of CFO and channel in OFDM relay systems is considered in \cite{Zhongshan2009, Thiagarajan2009}. In particular, a two-time-slot cooperative estimation protocol has been proposed in \cite{Zhongshan2009} for OFDM relay systems, while in \cite{Thiagarajan2009} the authors studied the maximum likelihood (ML) based, and the least squares based, joint CFO and channel estimation algorithms. However, none of the approaches in \cite{Zhongshan2009, Thiagarajan2009} consider the effect of PN on channel and CFO estimation or the overall relaying performance. While ignoring the effect of CFO, joint channel and PN estimation in OFDM relay networks is analyzed in \cite{Rabiei_tcom2011}. Although the approach in \cite{Rabiei_tcom2011} can be applied to AF relaying systems, it requires the relay to remove the cyclic prefix (CP) corresponding to the source-to-relay link and add a new CP before forwarding the OFDM symbol. Such an approach can result in significant additional overhead at the relay. Moreover, none of the approaches in \cite{Zhongshan2009, Thiagarajan2009,Rabiei_tcom2011} consider the effect of PN on joint channel and CFO estimation.


\vspace{-4pt}
\subsection{Contributions}

In this paper, different from \cite{Zhongshan2009, Thiagarajan2009}, the problem of joint CFO, PN, and channel estimation in OFDM relay systems is considered. Although joint CFO, PN, and channel estimation has been studied for point-to-point OFDM systems \cite{Jun2009, Darryl2006, Septier2008}, to the best of the authors' knowledge, this problem has not been considered in the context of relay systems. The contributions of this paper can be summarized as follows:
\begin{itemize}
\item A training and data transmission framework for OFDM relay networks is proposed that enables joint estimation of channel, CFO, and PN parameters at the destination.
\item A new \emph{hybrid Cramer-Rao lower bound (HCRLB)} for analyzing the performance of joint channel, CFO, and PN estimators in OFDM relay networks is derived.
\item An iterative joint channel, CFO, and PN estimator based on the \emph{maximum a posteriori (MAP)} criterion is proposed that exploits the correlation between PN parameters to significantly reduce estimation overhead.\footnote{The approach proposed here can be also applied to point-to-point systems to reduce PN estimation and carrier recovery overhead.} Moreover, the estimator's mean square error (MSE) performance is shown to be close to the derived HCRLB at moderate signal-to-noise ratios (SNRs).
\item A comb-type OFDM symbol containing both pilots and data symbols is proposed to track the time-varying PN parameters during the data transmission interval. Next, a novel iterative receiver that applies the proposed OFDM symbol to perform joint data detection and PN tracking at the destination node is derived.
\item Extensive simulations are carried out to investigate the performance of an OFDM relay system in the presence of CFO and PN. The results show that the combination of the proposed joint estimator and iterative receiver greatly enhances the bit error rate (BER) performance of OFDM relay systems with imperfect knowledge of channels, CFO, and PN.
\end{itemize}

\vspace{-4pt}
\subsection{Organization}
Section II presents the system model and assumptions in this paper. The joint estimation algorithm is presented in Section III. In Section IV, the HCRLB for the proposed joint estimation problem is derived. The proposed iterative receiver for joint data detection and PN tracking is present in Section V. Extensive simulation results are illustrated in Section VI. Finally, we conclude the paper in Section VII.

\vspace{-4pt}
\subsection{Notations}
Small italic letters, e.g., $x$ are for scalars, bold face small letters, e.g., $\mathbf{x}$, are used for vectors, and bold face capital alphabets, e.g., $\mathbf{X}$, are used for matrices. $\hat{x}$ is used to denote the estimate of $x$. $\mathbb{E}(\cdot)$ denotes the expectation of its argument. $\odot$, $\star$, and $\ast$ denote the Hadamard product, linear, and circular convolutions, respectively. ${\rm Tr}({\bf A})$, ${\bf A}^{-1}$, and $\det({\bf A})$ denote the trace, inverse, and determinant of matrix ${\bf A}$, respectively. ${\rm Diag}(\bf a)$ denotes a diagonal matrix with ${\bf a}$ being its diagonal entries. ${\rm Blkdiag}({\bf A}_0,{\bf A}_1,\cdots,{\bf A}_{N-1} )$ denotes a block diagonal matrix with ${\bf A}_0,{\bf A}_1,\cdots,{\bf A}_{N-1}$ as its diagonal matrices. ${\bf A}(N:M,:)$ and ${\bf A}(:,N:M)$ denote a submatrix containing the $N$-th to $M$-th rows of ${\bf A}$ and a submatrix containing the $N$-th to $M$-th columns of ${\bf A}$, respectively. Superscripts $(\cdot)^T$, $(\cdot)^{*}$ and $(\cdot)^H$ denote the transpose, conjugate, and conjugate transpose, respectively. ${\bf 0}_{N\times M}$, ${\bf I}_N$, and ${\bf 1}_N$ denote the $N\times M$ zero matrix, $N \times N$ identity matrix, and $N \times 1$ vector of ones, respectively. ${\Re}(z)$ and $\Im (z)$ denote the real and imaginary operators. ${\mathbb C}^{x \times y}$ and ${\mathbb R}^{x \times y}$ denote spaces of $x \times y$ matrices with complex and real entries, respectively. $\triangle^{{\bf x}}_{{\bf x}} f(\cdot) \triangleq  \frac{\partial f}{\partial {\bf x}}[\frac{\partial f}{\partial {\bf x}} f(\cdot)]^T$ denotes the second order partial derivative of function $f(\cdot)$ with respect to vector ${\bf x}$. Finally, ${\cal CN}({\bf x},{\bf \Sigma})$ and ${\cal N}({\bf x},{\bf \Sigma})$ denote real and complex Gaussian distributions, respectively, with mean $\bm \mu$ and covariance ${\bm \Sigma} $.

\vspace{-0pt}
\section{System Model}
\begin{figure}[t]
\begin{centering}
\includegraphics[scale=0.45]{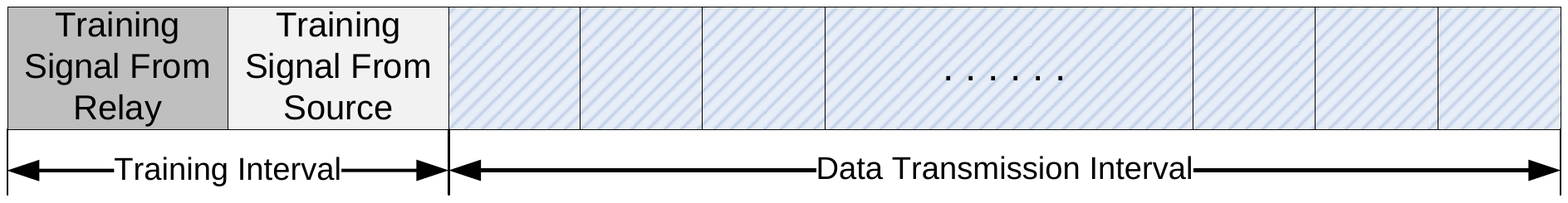}
\vspace{-0.2cm}
\caption{The proposed timing diagram of the OFDM relay system.
} \label{timing_diagram}
\end{centering}
\vspace{-0.5cm}
\end{figure}

An AF relaying OFDM system is considered, where a source node transmits its signal to a destination node through a relay. Unlike the work in \cite{Rabiei_tcom2011}, it is assumed that the relay node simply forwards the received signal without removing the CP corresponding to the source-to-relay link and appending a new CP for the relay-to-destination link. This assumption ensures a considerably simpler relaying structure. $N$ subcarriers are used for OFDM transmission. Similar to prior work in this field, e.g., \cite{Darryl2006,Darryl2007}, quasi static fading channels are considered, i.e., the CSI is assumed to be constant over the duration of a single packet. Each packet consists of two OFDM training symbols, which are followed by multiple data symbols as shown in Fig.~\ref{timing_diagram}. The two training symbols are used to separately estimate the channel responses and CFO in the presence of unknown PN for both the source to relay and relay to destination hops (Fig.~\ref{timing_diagram}).

The proposed signal model can be applied to both full-duplex and half-duplex relaying networks based on the following system setups and assumptions:
\begin{itemize}
\item \emph{Full-duplex relaying}: In this setup, the proposed signal model is applicable to relaying networks that utilize highly directional transmit and receive antennas with large antenna gains at the relay, e.g., microwave and millimeter-wave systems \cite{Wells-2010,book_E-band}. This approach minimizes or eliminates the effect of self-interference at the relay.\footnote{Application of sophisticated transceivers has also been shown to minimize or eliminate the impact of self-interference at the relay \cite{Youngki2010letter}.} Moreover, it is assumed that the relay forwards its signal to the destination in passband without converting it to baseband. This assumption is practical since there are various radio frequency (RF) amplifiers that can operate at high carrier frequencies and can be utilized in full-duplex relaying networks, e.g., Mini-Circuits AVA-$24$+ with a frequency range of $5$--$20$ GHz \cite{data_sheet}.

\item \emph{Half-duplex relaying}: In this setup, it is assumed that the relay forwards its received signal on a different carrier frequency and does not convert it to baseband, i.e., the relay applies on-frequency/on-channel RF relaying \cite{Salehian2002TBOC}. Moreover, the difference between the receive and transmit carrier frequencies are assumed to be small to enable the application of a low PN oscillator at the relay. An example of such an oscillator is ROS-$209$-$319$+ ultra low noise voltage controlled oscillator that has a very small PN factor of $-133$ dBc/Hz at an offset frequency of $10$ KHz \cite{data_sheet2}. As such, in this setup, it is assumed that the signal forwarded from the relay is not affected by PN.

\end{itemize}


\begin{figure}[t]
\begin{centering}
\includegraphics[scale=0.45]{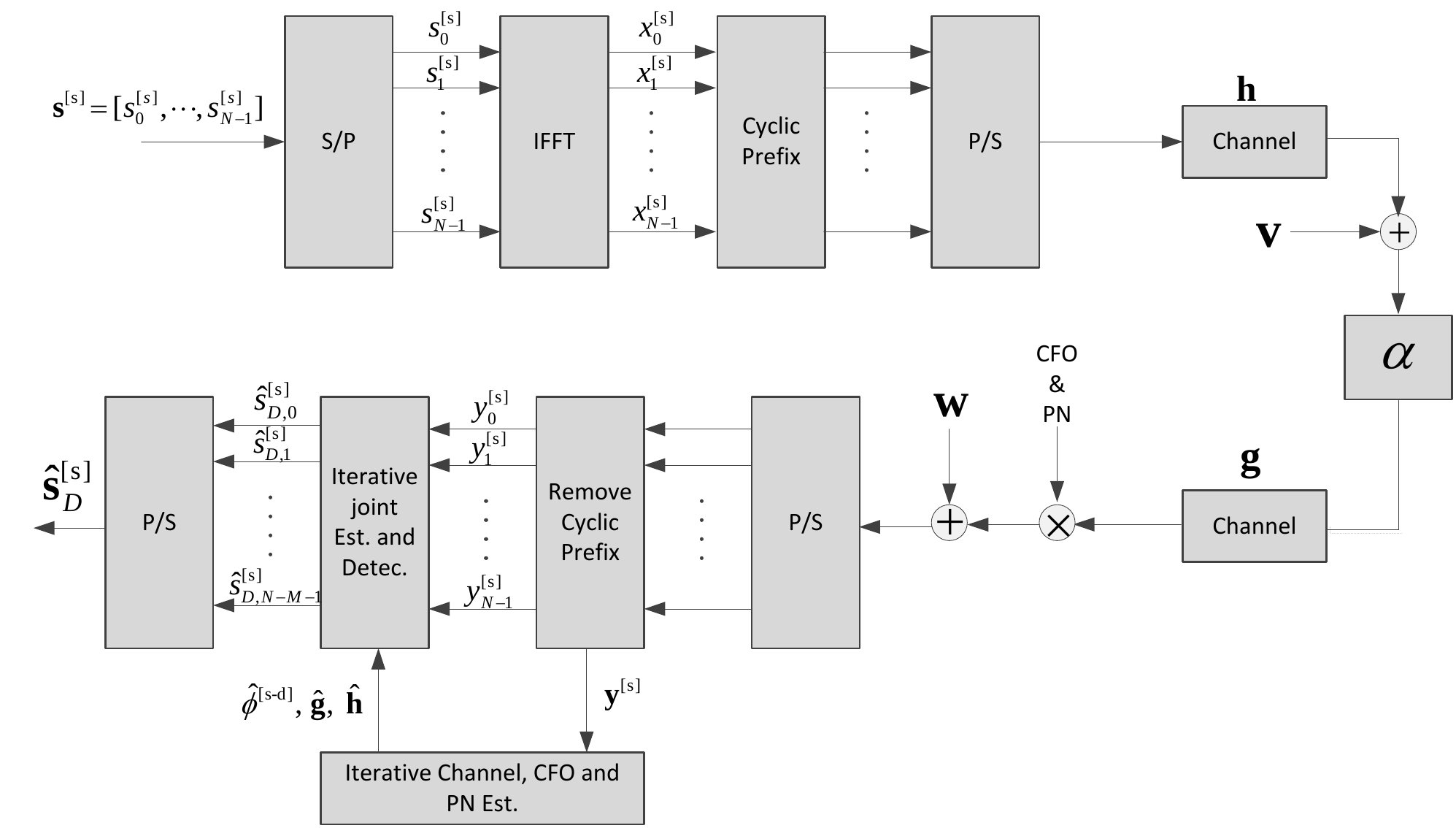}
\vspace{-0.3cm}
\caption{Illustration of transceiver structure of the OFDM relay system.} \label{Full_duplex}
\end{centering}
\vspace{-0.3cm}
\end{figure}

\vspace{-4pt}
\subsection{Signal Transmission from Source to Destination}

The overall transmission and reception structure of each OFDM symbol from the source to the destination node is illustrated in Fig.~\ref{Full_duplex}. Let ${\bf s}^{\text{[s]}}\triangleq \big[s^{\text{[s]}}_{0}, s^{\text{[s]}}_{1}, \cdots, s^{\text{[s]}}_{N-1} \big]^T$ denote the frequency domain modulated training or data signal sequence at the source node, which is then transformed into a set of parallel symbols $s^{\text{[s]}}_{k}$, for $k=0,\cdots,N-1$. By conducting an inverse fast Fourier transform (IFFT), we obtain the time domain signal vector ${\bf x}^{\text{[s]}}$ as ${\bf x}^{\text{[s]}} = {\bf F}^H {\bf s}^{\text{[s]}}$, where ${\bf x}^{\text{[s]}}\triangleq \big[x^{\text{[s]}}_{0}, x^{\text{[s]}}_{1}, \cdots, x^{\text{[s]}}_{N-1} \big]^T$, and ${\bf F}$ is the normalized discrete Fourier transform (DFT) matrix with ${\bf F}_{n,k} =\frac{1}{\sqrt{N}}\exp \big(-j\frac{2\pi (n-1) (k-1)}{N} \big)$. After adding the CP, the parallel signal vector is transformed into a time domain sequence denoted by $x^{\text{[s]}}(n)$, for $n=-L,\cdots, N-1$. Subsequently, the transmitted baseband continuous signal from the source, $\tilde{x}^{\text{[s]}}(t)$, can be written as
\begin{equation}\label{EQU-2}
\begin{split}
\hspace{-10pt}\tilde{x}^{\text{[s]}}(t) = \sum^{N-1}_{n=-L} x^{\text{[s]}}(n) q(t-nT_s),\;\;\;\;\;\;\;0\leq t \leq T+T_{CP}
\end{split}
\end{equation}
where $T_s = T/N$ with $T$ denoting an OFDM symbol duration, $q(t)$ is the pulse shaping filter, $T_{CP}$ is the duration of the CP, and $x^{\text{[s]}}(-n)=x^{\text{[s]}}(N-n)$, for $n=-L,\cdots, -1$, is the added CP symbol.

At the destination, the baseband received signal, $y^{\text{[s]}}(t)$, is given by
\vspace{-4pt}
\begin{equation}\label{EQU-4}
\begin{split}
y^{\text{[s]}}(t) = &\alpha e^{j \theta^{\text{[s-d]}}(t)} e^{j \phi^{\text{[s-d]}}(t)} \big[  g(t) \star h(t)\star \tilde{x}^{\text{[s]}}(t) \\
&+  g(t) \star  v(t)  \big] + w(t),
\end{split}
\end{equation}
where $\alpha$ is the constant and scalar amplification factor at the relay, $h(t)$ and ${g}(t)$ are the frequency-selective fading channels from source to relay and the relay to destination, respectively, and $v(t)$ and $w(t)$ are the additive noises at the relay and at the destination, respectively.
Using a similar approach to point-to-point systems \cite{Jun2009, Darryl2006, Septier2008}, in \eqref{EQU-4}, $\theta^{\text{[s-d]}}(t)$ is the PN corresponding to source-relay-destination link, while $\phi^{\text{[s-d]}}(t)\triangleq2\pi \triangle f^{\text{[s-d]}} t$ 
is the CFO caused by the unmatched source and destination carrier frequencies.

After sampling at a sampling rate of $1/T_s$ and removing the CP, the received signal at the destination is determined as
\begin{equation}\label{EQU-5}
\begin{split}
y^{\text{[s]}}(nT_s) =&  \alpha e^{j \theta^{\text{[s-d]}}(n T_s)} e^{j 2\pi \Delta f^{\text{[s-d]}} n T_s} \big[ \underbrace{g(n T_s) \star h(n T_s)}_{\triangleq c(n T_s)} \\
& \ast \tilde{x}^{\text{[s]}}(n T_s)  +   g(n T_s) \star  v(n T_s) \big]+ w(n T_s),\\
\end{split}
\end{equation}
where circular convolution appears in \eqref{EQU-5} due to the added CP at the source node.
Note that to avoid ICI, the length of CP, denoted by $N_{CP}=T_{CP}/T_s$, should be larger than $L=L_h + L_g -1$ with $L_h$ and $L_g$ being the number of channel taps of $h(t)$ and $g(t)$, respectively. Eq. \eqref{EQU-5} can be written in vector form as\footnote{For notational convenience, we discard the term $T_s$ in \eqref{EQU-9}.\vspace{-0pt}}\vspace{-0pt}
\begin{equation}\label{EQU-9}
\begin{split}
\mathbf{y}^{\text{[s]}} =\alpha {\bm \Lambda}_{\bm\theta^{\text{[s-d]}}} {\bm \Lambda}_{\phi^{\text{[s-d]}}} \left[ {\bf C} {\bf x}^{\text{[s]}} +  \mathbf{G} \mathbf{v} \right] + \mathbf{w},
\end{split}
\end{equation}
where\vspace{-0pt}
\begin{itemize}
\item $\mathbf{ y}^{\text{[s]}} \triangleq [y^{\text{[s]}}(0),y^{\text{[s]}}(1),\cdots,y^{\text{[s]}}(N-1)]^T$, 
\item ${\bm \Lambda}_{\bm\theta^{\text{[s-d]}}} \triangleq {\rm Diag}\big[ e^{j \theta^{\text{[s-d]}}(0)}, e^{j \theta^{\text{[s-d]}}(1)}, \cdots, e^{j \theta^{\text{[s-d]}}(N-1)} \big]$,
\item ${\bm \Lambda}_{\phi^{\text{[s-d]}}} \triangleq {\rm Diag}\big[ 1, e^{j 2\pi \phi^{\text{[s-d]}}/N }  , \cdots, e^{j 2\pi \phi^{\text{[s-d]}} (N-1)/N} \big]$, $\phi^{\text{[s-d]}} \triangleq {\Delta f}^{\text{[s-d]}} T$ is the normalized CFO,
\item ${\bf C} \triangleq {\bf F}^H {\bf \Lambda}_{\tilde{\bf c}} {\bf F}$, ${\bf \Lambda}_{\tilde{\bf c}}\triangleq{\rm Diag}(\tilde{{\bf c}}) $ with $\tilde{{\bf c}}\triangleq\sqrt{N}{\bf F} [{\bf c}^T, {\bf 0}^T_{N-L,1}]^T$ and ${\bf c}\triangleq[c(0),c(1),\cdots,c(L-1)]^T$,
\item ${\bf v}\triangleq [v(-L_g+1),\cdots, v(0),\cdots, v(N-1) ]^T$ and ${\bf w}_1 \triangleq [w(0),w(1),\cdots, w(N-1) ]^T$ are the sampled additive noise at the relay and destination nodes, respectively, and
\end{itemize}
\begin{equation}\label{EQU-10-1}
\begin{split}
{\bf G} & = \left[
                    \begin{array}{cccc}
                      g(L_g-1) & g(L_g-2) & \cdots & 0\\
                      0 & g(L_g-1) & \cdots  & 0  \\
                     \vdots & \vdots &  \ddots   & \vdots \\
                      0 & 0 & \cdots &g(0)\\
                    \end{array}
 \right]
 \end{split}
\end{equation}
is an $N \times (N+L_g-1)$ matrix.
The additive noise at the relay and destination are distributed as ${\bf v}\sim{\cal CN}(0, \sigma^2_R {\bf I}_{N+N_{CP}})$ and ${\bf w}\sim{\cal CN}(0, \sigma^2_D {\bf I}_N)$, respectively. 
Finally, although ${\bf C}$ is an $N \times N$ circulant matrix, ${\bf G}$ is a regular $N \times (N+L_g-1)$ matrix, since no CP is added at the relay node.

\subsection{Training Signal Transmission from Relay to Destination}

Recall that the second OFDM training symbol is transmitted from the relay to separately estimate the relay-to-destination channel. Following similar steps as above, the vector of received training signal at the destination node from the relay, ${\bf y}^{\text{[r]}}\triangleq[y^{\text{[r]}}(0),y^{\text{[r]}}(1),\cdots,y^{\text{[r]}}(N-1)]^T$, is given by\vspace{-0pt}
\begin{equation}\label{EQU-9-1}
\begin{split}
{\bf y}^{\text{[r]}} = {\bm \Lambda}_{\bm\theta^{\text{[r-d]}}} {\bm \Lambda}_{\phi^{\text{[r-d]}}}  \bar{{\bf G}} {\bf x}^{\text{[r]}} +  {\bf w},
\end{split}
\end{equation}
where\vspace{-0pt}
\begin{itemize}
\item ${\bf x}^{\text{[r]}} \triangleq[x^{\text{[r]}}(0), x^{\text{[r]}}(1), \cdots, x^{\text{[r]}}(N-1)]^T = {\bf F}^H {\bf s}^{\text{[r]}}$, ${\bf s}^{\text{[r]}}$ is the frequency domain relay training signal,
\item ${\bm \Lambda}_{\bm\theta^{\text{[r-d]}}} \triangleq {\rm Diag}\big[ e^{j \theta^{\text{[r-d]}}(0)}, e^{j \theta^{\text{[r-d]}}(1)}, \cdots, e^{j \theta^{\text{[r-d]}}(N-1)} \big]$, $\theta^{\text{[r-d]}}(n)$ is the $n$-th PN sample corresponding to relay-destination link,
\item ${\bm \Lambda}_{\phi^{\text{[r-d]}}} \triangleq {\rm Diag}\big[ 1, e^{j 2\pi \phi^{\text{[r-d]}}/N }  , \cdots, e^{j 2\pi \phi^{\text{[r-d]}} (N-1)/N} \big]$,
\item $\phi^{\text{[r-d]}}$ is the normalized CFO generated by the mismatch between the relay and destination carrier frequencies,
\item $\bar{{\bf G}}$ is a circulant channel matrix given by $\bar{{\bf G}} \triangleq {\bf F}^H {\bf \Lambda}_{\tilde{\bf g}} {\bf F}$, with ${\bf \Lambda}_{\bf g} = {\rm Diag}(\tilde{{\bf g}})$, $\tilde{{\bf g}}\triangleq\sqrt{N}{\bf F} [{\bf g}^T, {\bf 0}^T_{N-L_g,1}]^T$, and ${\bf g} \triangleq [g(0),g(1),\cdots,g(L_g-1)]^T$.
\end{itemize}
\subsection{Statistical Model of Phase Noise}\label{sec:short-PN}
Similar to \cite{Darryl2006} and based on the properties of PN in practical oscillators, PN is modeled by a Wiener process, i.e.,
\begin{equation}\label{EQU-10-1-1}
\begin{split}
\theta^{[i]}(n) = \theta^{[i]}(n-1) + \Delta^{[i]} (n),\;\;\;\;\;\;\;\;i=[\text{s-d}],[\text{r-d}]
\end{split}
\end{equation}
where $\Delta^{[i]} (n-1)$ is a real Gaussian variable following $\Delta^{[i]} (n) \thicksim {\cal N}(0,\sigma^2_{\Delta^{[i]} })$. Here $\sigma^2_{\Delta^{[i]} } = 2\pi \beta^{[i]} T_s$ with $\beta^{[i]}$ denoting the one-sided $3$-dB bandwidth of the Lorentzian spectrum of the oscillator
\cite{Chorti2006, Demir2000}.
As in \cite{Darryl2006,Darryl2007}, it is assumed that $\theta^{[i]}(-1) = 0$ since the residual PN at the start of the frame is estimated as part of the channel parameters. From \eqref{EQU-10-1-1}, it can be concluded that the PN vector, ${\bm \theta}^{[i]} \triangleq [\theta^{[i]}(0), \theta^{[i]}(1), \cdots, \theta^{[i]}(N-1)]^T$, follows a Gaussian distribution, i.e., ${\bm \theta}^{[i]} \thicksim {\cal N}(0,{\bf \Psi}^{[i]})$, where the covariance matrix ${\bf \Psi}^{[i]}$ is given by
\begin{equation}\label{III-10-2}
\begin{split}
{\bf \Psi}^{[i]} =\sigma^2_{\Delta^{[i]}} \left[
                                        \begin{array}{cccccc}
                                          1 & 1 & 1 & \cdots & 1 & 1 \\
                                          1 & 2 & 2 & \cdots & 2 & 2 \\
                                           1 & 2 & 3 & 3 & \cdots & 3 \\
                                          \vdots & \vdots & \vdots & \vdots & \ddots & \vdots \\
                                          1 & 2 & 3 & \cdots & N-1 & N
                                        \end{array}
                                      \right].
\end{split}
\end{equation}

\begin{figure}[t]
\begin{centering}
\vspace{-0.3cm}
\includegraphics[scale=0.5]{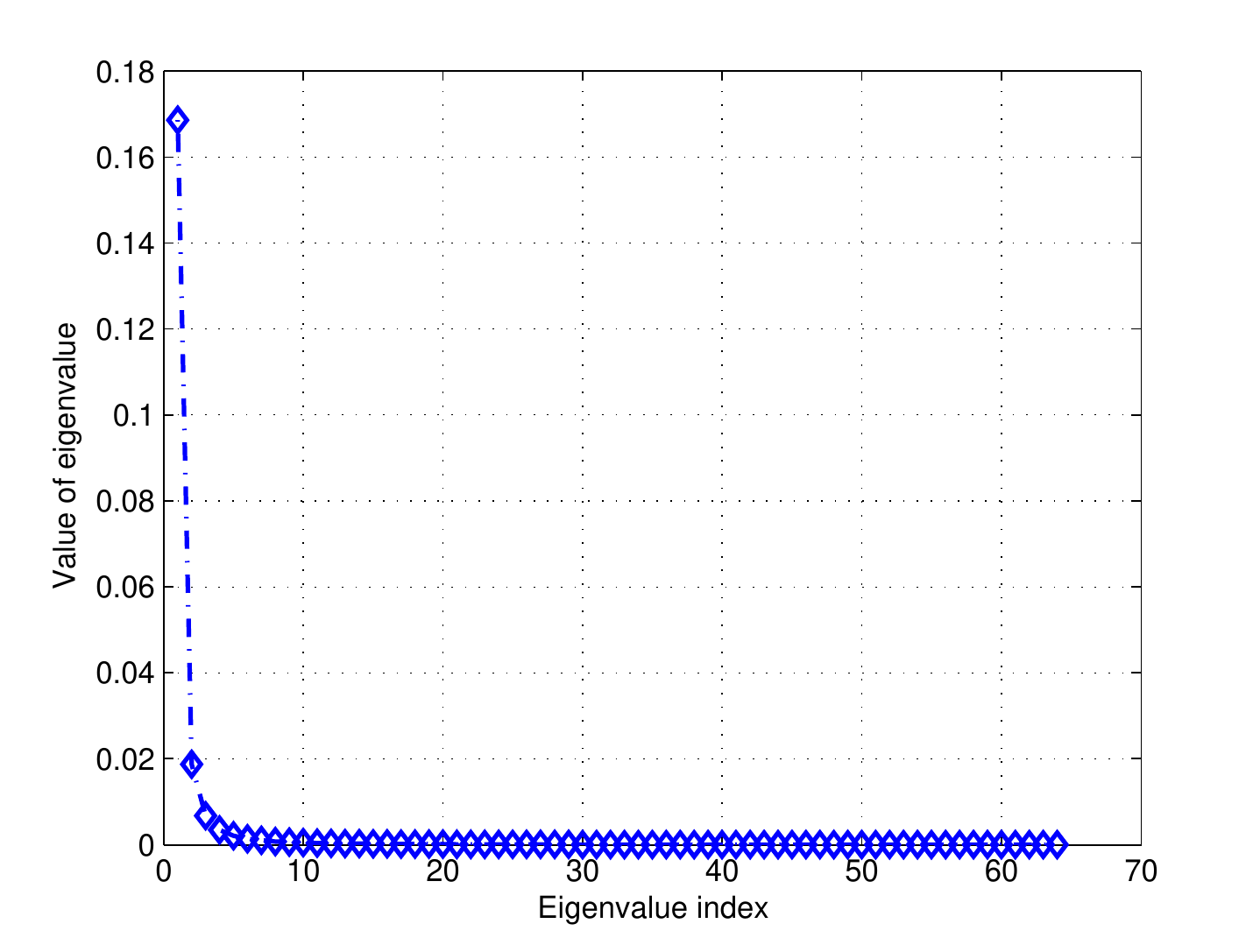}
\vspace{-0.4cm}
\caption{Illustration of the eigenvalues of ${\bf \Psi}^{[i]}$ at $N=64$ and $\sigma^2_{\Delta^{[i]} } = 10^{-4}$.} \label{Eigen}
\end{centering}
\vspace{-0.3cm}
\end{figure}
In obtaining the covariance matrix in \eqref{III-10-2}, similar to prior results in this field \cite{Darryl2007}, it is assumed that the PN variances are small enough such that $\theta^{[i]}(n)$ does not reach its maximum value of $\pi$. This assumption is justifiable since practical oscillators have a very small PN variance as shown in \cite{Mehrpouyan2012}.

%
%

Based on the signal model in \eqref{EQU-9} and \eqref{EQU-9-1}, it can be observed that a large number of channel, CFO, and PN parameters need to be jointly estimated, which increases the computational complexity of the receiver at the destination. Thus, to reduce estimation overhead, we take advantage of the correlation amongst the PN parameters to reduce the number of unknown parameters that need to be estimated. The eigenvalues of the covariance matrix, ${\bf \Psi}^{[i]}$ are illustrated in Fig.~\ref{Eigen}. From this figure it can be deduced that most eigenvalues of the matrix ${\bf \Psi}^{[i]}$ are close to zero. Thus, the PN vector, ${\bm \theta}^{[i]}$, can be represented as
\vspace{-6pt}
\begin{equation}\label{EQU-10-3}
\begin{split}
{\bm \theta}^{[i]} = {\bf \Pi}^{[i]} {\bm \eta}^{[i]}, \;\;\;\;~i=[\text{s-d}],[\text{r-d}]
\end{split}
\end{equation}
where ${\bm \eta}^{[i]} \thicksim {\cal N}(0,{\bf I}_M)\in \mathbb{C}^{M \times 1}$ is the shortened unknown PN vector containing $M\leq N$ PN parameters, while ${\bf \Pi}^{[i]} \in \mathbb{C}^{N \times M}$ is the transformation matrix that allows for obtaining ${\bm \eta}^{[i]}$ from ${\bm \theta}^{[i]}$. Moreover, the singular value decomposition of ${\bf \Psi}^{[i]}$ is given by
${\bf \Psi}^{[i]} = {\bf U}^{[i]} {\bf D}^{[i]} \left({\bf U}^{[i]}\right)^T$, where ${\bf U}^{[i]}$ is the $N \times N$ eigenvector matrix of ${\bf \Psi}^{[i]}$ and ${\bf D}^{[i]} = {\rm Diag}({\bm \nu}^{[i]})$. Here, ${\bm \nu}^{[i]}\triangleq[\nu_{i,0}, \nu_{i,1}, \cdots, \nu_{i,N-1} ]^T$ is the vector of the eigenvalues of ${\bf \Psi}^{[i]}$ arranged in decreasing order.
Subsequently, the matrix ${\bf \Pi}^{[i]}$ in \eqref{EQU-10-3} can be selected as ${\bf \Pi}^{[i]} = \tilde{{\bf U}}^{[i]} \tilde{{\bf D}}^{[i]}$, where $\tilde{{\bf U}}^{[i]} = {\bf U}^{[i]}(:, 0:M-1)$ and $\tilde{{\bf D}}^{[i]} = {\rm Diag}(\tilde{{\bm \nu}}^{[i]})$ with $\tilde{{\bm \nu}}^{[i]}\triangleq\big[\sqrt{\nu_{i,0}}, \sqrt{\nu_{i,1}}, \cdots, \sqrt{\nu_{i,M-1}} \big]^T$. In the subsequent sections, ${\bm \eta}^{[i]}$, for $i=[\text{s-d}],[\text{r-d}]$ is estimated instead of ${\bm \theta}^{[i]}$. A suitable choice of $M$ that allows for accurate PN tracking is presented in Section \ref{sec:simulations}.

\vspace{-4pt}
\section{Proposed Joint Channel, CFO and Phase Noise Estimation}

In order to avoid the negative impact of ICI caused by CFO and PN, in this work, the joint estimation of channel parameters, CFO, and PN is performed in the time domain. To proceed, we reformulate \eqref{EQU-9} as \vspace{-0pt}
\begin{equation}\label{III-1}
\begin{split}
{\bf y}^{\text{[s]}}
=& \alpha {\bm \Lambda}_{\bm\theta^{\text{[s-d]}}} {\bm \Lambda}_{\phi^{\text{[s-d]}}} \left( {\bf F}^H {\bm \Lambda}_{s^{\text{[s]}}} {\bf F}_{[L]}{\bf c} +  {\bf G} {\bf v} \right) + {\bf w},
\end{split}
\end{equation}
where ${\bf s}^{\text{[s]}}$ denotes the training symbol transmitted from source such that $\mathbb{E}\Big({\bf s}^{\text{[s]}} \left[{\bf s}^{\text{[s]}}\right]^H\Big) = P^{\text{[s]}}_{\text{T}}{\bf I}_N$, $P^{\text{[s]}}_{\text{T}}$ is the transmit power from the source, ${\bm \Lambda}_{s^{\text{[s]}}}\triangleq{\rm Diag} ({\bf s}^{\text{[s]}})$, and ${\bf F}_{[L]}\triangleq\sqrt{N}{\bf F}(:,0:L-1)$. Similarly, the received signal ${\bf y}^{\text{[r]}}$ in \eqref{EQU-9-1} can be rewritten as\vspace{-8pt}
\begin{equation}\label{III-1-1}
\begin{split}
{\bf y}^{\text{[r]}} & = {\bm \Lambda}_{\bm\theta^{\text{[r-d]}}} {\bm \Lambda}_{\phi^{\text{[r-d]}}}  {\bf F}^H {\bm \Lambda}_{s^{\text{[r]}}} {\bf F}_{[L_g]}{\bf g} + {\bf w},
\end{split}
\end{equation}
where ${\bm \Lambda}_{s^{\text{[r]}}}\triangleq{\rm Diag} ({\bf s}^{\text{[r]}})$ denotes the training symbol transmitted from relay such that $\mathbb{E}({\bf s}^{\text{[r]}} \left[{\bf s}^{\text{[r]}}\right]^H) = P^{\text{[r]}}_{\text{T}}{\bf I}_N$, $P^{\text{[r]}}_{\text{T}}$ is the transmit power from the relay, and ${\bf F}_{[L_g]}\triangleq\sqrt{N}{\bf F}(:,0:L_g-1)$. As in \cite{Darryl2006}, it is assumed that ${\bf s}^{\text{[s]}}$ and ${\bf s}^{\text{[r]}}$ are known constant-modulus training symbols.

From the detection point of view, it may appear that one only needs to estimate the CFO, $\phi^{\text{[s-d]}}$, and the \emph{combined} source-relay-destination channel, ${\bf c}$, in the presence of PN, ${\bm \theta}^{\text{[s-d]}}$. However, as shown in \eqref{III-1}, the relay-to-destination channel, ${\bf g}$, affects the statistic of the additive noise at the destination node. Hence, it needs to be known to develop a joint PN estimation and data detection algorithm based on the MAP criterion. Consequently, here, the parameters of interest are: the CFO, $\phi^{\text{[s-d]}}$, the channel from source to relay, ${\bf h} \triangleq [h(0),h(1),\cdots,h(L_h-1)]^T$ and the relay to destination channel, ${\bf g}$. Moreover, in addition to the parameters of interest, there are also unknown nuisance parameters, e.g., the CFO and PN from relay to destination, $\phi^{\text{[r-d]}}$ and ${\bm \theta}^{\text{[r-d]}}$, respectively, that also need to be jointly estimated. Using the approach in \cite{Darryl2006} and the received training signal from the relay node, ${\bf y}^{\text{[r]}}$, the MAP estimates of the CFO from relay to destination, $\phi^{\text{[r-d]}}$, can be obtained as
\begin{align}\label{III-2}
\hat{\phi}^{\text{[r-d]}}  = & \arg \min_{\phi^{\text{[r-d]}}} - {\bf 1}^T_N \Im ({\bm \Lambda}_{\phi^{\text{[r-d]}}} {\bf A}  {\bf A}^H {\bm \Lambda}^H_{\phi^{\text{[r-d]}}} )^T\\
&\times \Big[\Re({\bm \Lambda}_{\phi^{\text{[r-d]}}} {\bf A}  {\bf A}^H {\bm \Lambda}^H_{\phi^{\text{[r-d]}}}) + \frac{\sigma^2 P^{\text{[r]}}_{^{\text{T}}}}{2} [{\bf \Psi}^{\text{[r-d]}}]^{-1} \Big]^{-1}  \notag\\
&\times
\Im ({\bm \Lambda}_{\phi^{\text{[r-d]}}} {\bf A}  {\bf A}^H {\bm \Lambda}^H_{\phi^{\text{[r-d]}}} ){\bf 1}_N
 + {\bf 1}^T_N {\bm \Lambda}_{\phi^{\text{[r-d]}}} {\bf A}  {\bf A}^H {\bm \Lambda}^H_{\phi^{\text{[r-d]}}} {\bf 1}_N,\notag
\end{align}
where ${\bf A} \triangleq  [{\bf Y}^{\text{[r]}}]^H {\bf F}^H {\bm \Lambda}_{s^{\text{[r]}}} {\bf V}$, ${\bf Y}^{\text{[r]}} \triangleq {\rm Diag}({\bf y}^{\text{[r]}})$, and ${\bf V} \triangleq {\bf F}(:,{L_g:N-1})$. Using the estimated CFO from relay to destination, $\hat{\phi}^{\text{[r-d]}}$, the PN vector ${\bm \theta}^{\text{[r-d]}}$ is estimated as
\begin{align}\label{III-3}
\hat{\bm \theta}^{\text{[r-d]}} =& {\bf \Pi}^{\text{[r-d]}}\Big[\big[{\bf \Pi}^{\text{[r-d]}}\big]^T\Re\big(\hat{{\bm \Lambda}}_{\phi^{\text{[r-d]}}} {\bf A}  {\bf A}^H \hat{{\bm \Lambda}}^H_{\phi^{\text{[r-d]}}}\big){\bf \Pi}^{\text{[r-d]}} \\
& + \frac{\sigma^2 P^{\text{[r]}}_{^{\text{T}}}}{2} {\bf I}_M \Big]^{-1}  \big[{\bf \Pi}^{\text{[r-d]}}\big]^T \Im \big(\hat{{\bm \Lambda}}_{\phi^{\text{[r-d]}}} {\bf A}  {\bf A}^H \hat{{\bm \Lambda}}^H_{\phi^{\text{[r-d]}}} \big){\bf 1}_N,\notag
\end{align}
where $[\hat{{\bm \Lambda}}_{\phi^{\text{[r-d]}}}]_{m,m}  = \exp\big(\frac{j2\pi (m-1)\hat{\phi}^{\text{[r-d]}}}{N}\big)$. Unlike, the approach in \cite{Darryl2006}, in \eqref{III-3}, the shortened PN vector ${\bm \eta}^{\text{[r-d]}}$ is estimated first which reduces the complexity of the estimator by requiring the calculation of a considerably smaller matrix inverse. Based on the estimated $\hat{\phi}^{\text{[r-d]}}$ and $\hat{\bm \theta}^{\text{[r-d]}}$, the remaining parameters of interest can be estimated via ${\bf y}^{\text{[s]}}$ and ${\bf y}^{\text{[r]}}$.

From \eqref{III-1} and \eqref{III-1-1}, it can be observed that the joint estimation of channel response, CFO, and PN is a hybrid estimation problem consisting of both deterministic parameters, $\phi^{\text{[s-d]}}$, ${\bf h}$, ${\bf g}$, and random parameters, ${\bm \eta}^{\text{[s-d]}}$.
The posterior distribution of the parameters of interests given the received signals, ${\bf y}^{\text{[s]}}$ and ${\bf y}^{\text{[r]}}$, can be written as\vspace{-0pt}
\begin{align}\label{III-4}
\hspace{-10pt} p\big(\phi^{\text{[s-d]}}, {\bm \eta}^{\text{[s-d]}}, {\bf h}, {\bf g}| {\bf y}^{\text{[s]}}, {\bf y}^{\text{[r]}}\big) =& 
p\big({\bf y}^{\text{[s]}}, {\bf y}^{\text{[r]}}| \phi^{\text{[s-d]}}, {\bm \eta}^{\text{[s-d]}}, {\bf h}, {\bf g}\big) \notag\\
&\times
p\big({\bm \eta}^{\text{[s-d]}}\big)/ p\big({\bf y}^{\text{[s]}}, {\bf y}^{\text{[r]}}\big).
\end{align}
Maximizing the posterior distribution in
\eqref{III-4} is equivalent to minimizing the negative log-likelihood function (LLF) $\mathcal{L}(\phi^{\text{[s-d]}}, {\bm \eta}^{\text{[s-d]}}, {\bf h}, {\bf g}) = -\log \left(p({\bf y}^{\text{[s]}}, {\bf y}^{\text{[r]}}| \phi^{\text{[s-d]}}, {\bm \eta}^{\text{[s-d]}}, {\bf h}, {\bf g})\right) - \log (p({\bm \eta}^{\text{[s-d]}}))$. Our objective is to find the joint estimates of $\phi^{\text{[s-d]}}$, ${\bm \eta}^{\text{[s-d]}}$, ${\bf h}$, and ${\bf g}$ by optimizing the following unconstrained function
\begin{align}\label{III-6}
\hspace{-10pt}\{\hat{\phi}^{\text{[s-d]}}, \hat{{\bm \eta}}^{\text{[s-d]}}, \hat{{\bf h}}, \hat{{\bf g}}\} \propto  \arg \min_{\phi^{\text{[s-d]}}, {\bm \eta}^{\text{[s-d]}}, {\bf h}, {\bf g}} \mathcal{L}(\phi^{\text{[s-d]}}, {\bm \eta}^{\text{[s-d]}}, {\bf h}, {\bf g})
\end{align}
where $\mathcal{L}(\phi^{\text{[s-d]}}, {\bm \eta}^{\text{[s-d]}}, {\bf h}, {\bf g}) = \log \det({\bf \Sigma})+({\bf y} - {\bm \mu})^H {\bf \Sigma}^{-1} ({\bf y} - {\bm \mu}) + \frac{1}{2}[{\bm \eta}^{\text{[s-d]}}]^T {\bm \eta}^{\text{[s-d]}} $, ${\bf y} \triangleq \left[ [{\bf y}^{\text{[s]}}]^T, [{\bf y}^{\text{[r]}}]^T \right]^T$, ${\bm \mu} \triangleq \Big[(\alpha {\bm \Lambda}_{\bm\theta^{\text{[s-d]}}} $ ${\bm \Lambda}_{\phi^{\text{[s-d]}}}  {\bf F}^H {\bm \Lambda}_{s^{\text{[s]}}} {\bf F}_{[L]}{\bf c})^T, ({\bm \Lambda}_{\bm\theta^{\text{[r-d]}}} {\bm \Lambda}_{\phi^{\text{[r-d]}}}  {\bf F}^H {\bm \Lambda}_{s^{\text{[r]}}} {\bf F}_{[L_g]}{\bf g})^T \Big]^T $, ${\bf \Sigma} \triangleq{\rm Blkdiag} \left({\bf \Sigma}^{\text{[r]}}, {\bf \Sigma}^{\text{[d]}}\right)$, ${\bf \Sigma}^{\text{[r]}} = \alpha^2 \sigma^2_R {\bm \Lambda}_{\bm\theta^{\text{[s-d]}}} {\bm \Lambda}_{\phi^{\text{[s-d]}}}$ ${\bf G}{\bf G}^H  {\bm \Lambda}^H_{\phi^{\text{[s-d]}}} {\bm \Lambda}^H_{\bm\theta^{\text{[s-d]}}}+ \sigma^2_D {\bf I}_N$, and ${\bf \Sigma}^{\text{[d]}} = \sigma^2_D {\bf I}_N$. Although the CFO, $\phi^{\text{[s-d]}}$, and PN vector, ${\bm \theta}^{\text{[s-d]}}$, are only contained in the received signal, ${\bf y}^{\text{[s]}}$, the backward substitution method proposed in \cite{Darryl2006} cannot be exploited here to solve \eqref{III-6} due to the unknown noise covariance matrix ${\bf \Sigma}^{\text{[r]}}$. Moreover, since all the parameters of interest are coupled with each other, the optimization problem in \eqref{III-6} is a non-convex problem. To make \eqref{III-6} tractable, in the following subsections, we propose to decouple \eqref{III-6} into several subproblems that can be each solved separately in an iterative approach.

\subsection{Phase Noise Estimation}\label{sec:PN}
In the first subproblem, we intend to obtain an estimate of the PN vector ${\bm \eta}^{\text{[s-d]}}$ at the $(k+1)$-th iteration, $[\hat{{\bm \eta}}^{\text{[s-d]}}]^{[k+1]}$, via the estimates of $[{\bm \phi}^{\text{[s-d]}}]$, ${\bf h}$, and $\bf g$ from the $k$-th iteration, $[\hat{{\bm \phi}}^{\text{[s-d]}}]^{[k]}$, $\hat{\bf h}^{[k]}$ and $\hat{\bf g}^{[k]}$, respectively, according to
\begin{equation}\label{III-7}
\begin{split}
[\hat{{\bm \eta}}^{\text{[s-d]}}]^{[k+1]} \varpropto & \arg \min_{{\bm \eta}^{\text{[s-d]}}}~ \mathcal{L}_{{\bm \eta}^{\text{[s-d]}}}
\end{split}
\end{equation}
where $\mathcal{L}_{{\bm \eta}^{\text{[s-d]}}} =\log \det({\bf \Sigma}^{\text{[r]}})+({\bf y}^{\text{[s]}} - {\bm \mu}^{\text{[s-d]}})^H [{\bf \Sigma}^{\text{[r]}}]^{-1} ({\bf y}^{\text{[s]}} - $\\${\bm \mu}^{\text{[s-d]}}) + \frac{1}{2}[{\bm \eta}^{\text{[s-d]}}]^T {\bm \eta}^{\text{[s-d]}}$ with
${\bm \mu}^{\text{[s-d]}} \triangleq \alpha {\bm \Lambda}_{\bm\theta^{\text{[s-d]}}} \hat{{\bm \Lambda}}^{[k]}_{\phi^{\text{[s-d]}}}  {\bf F}^H {\bm \Lambda}_{s^{\text{[s]}}} {\bf F}_{[L]} $\\$\hat{{\bf c}}^{[k]}$, $[\hat{{\bm \Lambda}}^{[k]}_{\phi^{\text{[s-d]}}}]_{m,m} = \exp(\frac{j2\pi (m-1)[\hat{\phi}^{\text{[s-d]}}]^{[k]}}{N})$, $\hat{{\bf c}}^{[k]} \triangleq \hat{\bf h}^{[k]} \star \hat{\bf g}^{[k]}$,
${\bf \Sigma}^{\text{[r]}} = \alpha^2 \sigma^2_R {\bm \Lambda}_{\bm\theta^{\text{[s-d]}}} \hat{{\bm \Lambda}}^{[k]}_{\phi^{\text{[s-d]}}} \hat{\bf G}^{[k]} [\hat{\bf G}^{[k]}]^H  [\hat{{\bm \Lambda}}^{[k]}_{\phi^{\text{[s-d]}}}]^H {\bm \Lambda}^H_{\bm\theta^{\text{[s-d]}}} + \sigma^2_D {\bf I}_N$, and $\hat{\bf G}^{[k]}$ is constructed from $\hat{\bf g}^{[k]}$ as shown in \eqref{EQU-10-1}.
As shown in Appendix~\ref{prof_PN}, a closed-form solution for the PN estimate at the $(k+1)$-th iteration, $[{\bm \eta}^{\text{[s-d]}}]^{[k+1]}$, can be found as
\begin{equation}\label{III-10}
\begin{split}
[\hat{{\bm \eta}}^{\text{[s-d]}}]^{[k+1]} =& \big[\Re({\bf B}^H \left[[\hat{{\bf \Sigma}}^{\text{[r]}}]^{[k]} \right]^{-1} {\bf B}) + \frac{1}{2}{\bf I}_M \big]^{-1} \\
& \times \Re({\bf B}^H \left[[\hat{{\bf \Sigma}}^{\text{[r]}}]^{[k]} \right]^{-1} \bar{{\bf y}}^{\text{[s]}} ),
\end{split}
\end{equation}
where $[\hat{{\bf \Sigma}}^{\text{[r]}}]^{[k]}$ is the estimate of the noise covariance matrix at the $k$-th iteration. Using \eqref{III-10}, the un-shortened PN estimates at the $(k+1)$-th iteration, $[\hat{{\bm \theta}}^{\text{[s-d]}}]^{[k+1]}$, can be determined as $[\hat{{\bm \theta}}^{\text{[s-d]}}]^{[k+1]} = {\bf \Pi}^{\text{[s-d]}} [\hat{{\bm \eta}}^{\text{[s-d]}}]^{[k+1]}$ (see Section \ref{sec:short-PN}). Finally, the noise covariance matrix, $[\hat{{\bf \Sigma}}^{\text{[r]}}]^{[k]}$, is updated via $[\hat{{\bm \theta}}^{\text{[s-d]}}]^{[k+1]}$.

\vspace{-6pt}
\subsection{Relay to Destination Channel Estimation}
In the second subproblem, the channel response ${\bf g}$ is updated by applying the estimated CFO, source-to-relay channel, and PN vector, $[\hat{\phi}^{\text{[s-d]}}]^{[k]}$, $\hat{\bf h}^{[k]}$ and $[\hat{\bm \theta}^{\text{[s-d]}}]^{[k+1]}$, respectively. To proceed, the combined channel ${\bf c}$ is first rewritten as \vspace{-0pt}
\begin{equation}\label{III-11}
\begin{split}
{\bf c} = \tilde{{\bf G}} {\bf h} = \tilde{{\bf H}} {\bf g},
\end{split}
\end{equation}
where $\tilde{{\bf G}} \in \mathbb{C}^{L \times L_h} $ is denoted as
\begin{equation}\label{III-12}
\begin{split}
\tilde{{\bf G}} = \left[
                    \begin{array}{ccccc}
                      g(0) & 0 & \cdots  & 0 & 0 \\
                      g(1) & g(0) & \cdots & 0 & 0 \\
                      \vdots & \vdots & \vdots &  \vdots & \vdots \\
                      0 & 0 & \cdots &  g(0) & 0 \\
                      0 & 0 & \cdots &  g(1) & g(0) \\
                    \end{array}
                  \right],
\end{split}
\end{equation}
and $\tilde{{\bf H}} \in \mathbb{C}^{L \times L_g} $ has a similar form as $\tilde{{\bf G}}$.
%
Subsequently, the optimization problem for updating the relay-to-destination channel, ${\bf g}$, is given by\vspace{-0pt}
\begin{align}\label{III-14}
\hat{{\bf g}}^{[k+1]}  \propto& \arg \min_{{\bf g}} \mathcal{L}_{\bf g} \\
\propto&\arg \min_{{\bf g}} \log \det({\bf \Sigma})+({\bf y} - {\bf C}{\bf g})^H {\bf \Sigma}^{-1} ({\bf y} - {\bf C}{\bf g}),\notag
\end{align}
where ${\bf C} \triangleq \Big[ \big(\alpha \hat{{\bm \Lambda}}^{[k+1]}_{{\bm\theta}^{\text{[s-d]}}} \hat{{\bm \Lambda}}^{[k]}_{\phi^{\text{[s-d]}}}  {\bf F}^H {\bm \Lambda}_{s^{\text{[s]}}} {\bf F}_{[L]}\hat{\tilde{{\bf H}}}^{[k]}\big)^T, \big(\hat{\bm \Lambda}_{\bm\theta^{\text{[r-d]}}} \hat{\bm \Lambda}_{\bm\phi^{\text{[r-d]}}}$  ${\bf F}^H {\bm \Lambda}_{s^{\text{[r]}}} {\bf F}_{[L_g]}\big)^T \Big]^T$ with $\hat{\tilde{{\bf H}}}^{[k]}$ being formed by using the estimate of the source-to-relay channel in the $k$-th iteration $\hat{\bf h}^{[k]}$ according to \eqref{III-11}, and ${\bf \Sigma} \triangleq {\rm Blkdiag}  \left({\bf \Sigma}^{\text{[r]}}, \sigma^2_D {\bf I}_N\right)$ with ${\bf \Sigma}^{\text{[r]}} = \alpha^2 \sigma^2_R \hat{{\bm \Lambda}}^{[k+1]}_{{\bm\theta}^{\text{[s-d]}}} \hat{{\bm \Lambda}}^{[k]}_{\phi^{\text{[s-d]}}} {\bf G}{\bf G}^H  [\hat{{\bm \Lambda}}^{[k]}_{\phi^{\text{[s-d]}}}]^H [\hat{{\bm \Lambda}}^{[k+1]}_{{\bm\theta}^{\text{[s-d]}}}]^H+ \sigma^2_D {\bf I}_N$. Since the covariance matrix ${\bf \Sigma}$ is dependent on the channel response ${\bf g}$ as shown in \eqref{III-6}, it is impossible to find a closed-form solution for ${\bf g}$ based on \eqref{III-14}. Thus, we propose to use the channel covariance matrix at the $k$-th (previous iteration), $\hat{{\bf \Sigma}}^{[k]}$, to obtain an estimate of ${\bf g}$ at the $(k+1)$-th iteration. Using this approach and by equating the gradient of $\mathcal{L}_{\bf g}$ in \eqref{III-14} to zero, a closed-form solution for the relay-to-destination channel at the $(k+1)$-th iteration, $\hat{{\bf g}}^{[k+1]}$, can be derived as
\vspace{-4pt}
\begin{equation}\label{III-16}
\begin{split}
\hat{\bf g}^{[k+1]} = \big({\bf C}^H \big[\hat{{\bf \Sigma}}^{[k]}\big]^{-1} {\bf C} \big)^{-1} {\bf C}^H [\hat{{\bf \Sigma}}^{[k]}]^{-1} {\bf y}.
\end{split}
\end{equation}
Subsequently, using $\hat{\bf g}^{[k+1]}$, the noise covariance $[\hat{{\bf \Sigma}}^{\text{[r]}}]^{[k]}$ is updated.

\vspace{-10pt}
\subsection{Source to Relay Channel Estimation}
In the third subproblem, we intend to update the estimate of the source to relay channel based on the estimates  $[\phi^{\text{[s-d]}}]^{[k]}$, $[{\bm \theta}^{\text{[s-d]}}]^{[k+1]}$, and ${\bf g}^{[k+1]}$ via the following optimization problem
\begin{align}\label{III-17}
 \hat{{\bf h}}^{[k+1]} \propto& \arg \min_{{\bf h}} \mathcal{L}_{\bf h} \\
 \propto& \arg \min_{{\bf g}} ~({\bf y}^{\text{[s]}} - {\bf D}{\bf h})^H \left[[\hat{{\bf \Sigma}}^{\text{[r]}}]^{[k]} \right]^{-1} ({\bf y}^{\text{[s]}} - {\bf D}{\bf h}),\notag
\end{align}
where ${\bf D}\triangleq \alpha \hat{{\bm \Lambda}}^{[k+1]}_{{\bm \theta}^{\text{[s-d]}}} \hat{{\bm \Lambda}}^{[k]}_{\phi^{\text{[s-d]}}}  {\bf F}^H {\bm \Lambda}_{s^{\text{[s]}}} {\bf F}_{[L]}\hat{\tilde{{\bf G}}}^{[k+1]}$. In \eqref{III-17}, $\hat{\tilde{{\bf G}}}^{[k+1]}$ is formed as indicated in \eqref{III-12} by using $\hat{\bf g}^{[k+1]}$. Similar to the relay to destination channel, ${\bf g}$, the closed-form solution of ${\bf h}$ in \eqref{III-17} can be obtained as
\begin{equation}\label{III-18}
\begin{split}
\hat{\bf h}^{[k+1]} = \big({\bf D}^H \big[[\hat{{\bf \Sigma}}^{\text{[r]}}]^{[k]} \big]^{-1} {\bf D} \big)^{-1} {\bf D}^H \big[[\hat{{\bf \Sigma}}^{\text{[r]}}]^{[k]} \big]^{-1} {\bf y}^{\text{[s]}}.
\end{split}
\end{equation}

\vspace{-6pt}
\subsection{CFO Estimation}\label{sec:CFO}
In order to find an estimate of the source-destination CFO at the $(k+1)$-th iteration, $[\hat{\phi}^{\text{[s-d]}}]^{[k+1]}$, similar to the steps in \eqref{III-7}, we approximate the covariance matrix, ${\bf \Sigma}^{\text{[r]}}$ with $[\hat{{\bf \Sigma}}^{\text{[r]}}]^{[k]}$ and solve the unconstrained problem
\begin{align}\label{III-18-1}
 [\hat{\phi}^{\text{[s-d]}}]^{[k+1]} \propto& \arg \min_{{\phi}^{\text{[s-d]}}} ~\mathcal{L}_{{\phi}^{\text{[s-d]}}} \notag\\
 \propto& \arg \min_{{\phi}^{\text{[s-d]}}}~ ({\bf y}^{\text{[s]}} - {\bm \mu}_{{\phi}^{\text{[s-d]}}})^H \big[[\hat{{\bf \Sigma}}^{\text{[r]}}]^{[k]} \big]^{-1} \notag\\
 &\times
 ({\bf y}^{\text{[s]}} - {\bm \mu}_{{\phi}^{\text{[s-d]}}}),
\end{align}
where ${\bm \mu}_{{\phi}^{\text{[s-d]}}} \triangleq \alpha {\bm \Lambda}_{\bm\phi^{\text{[s-d]}}} \hat{{\bm \Lambda}}^{[k+1]}_{{\bm\theta}^{\text{[s-d]}}}{\bf F}^H {\bm \Lambda}_{s^{\text{[s]}}} {\bf F}_{[L]}\hat{{\bf c}}^{[k+1]}$. To make the problem in \eqref{III-18-1} more tractable and find a closed-form solution, a Taylor series approximation similar to that in \eqref{III-7} is applied here. Accordingly, $e^{\frac{j2\pi m \phi^{\text{[s-d]}} }{N}}$ can be approximated as
\begin{equation}\label{III-19}
\begin{split}
 e^{\frac{j2\pi m \phi^{\text{[s-d]}} }{N}} \approx & e^{\frac{j2\pi m [\hat{\phi}^{\text{[s-d]}}]^{[k]} }{N}}  \\
 &+ \big({\phi}^{\text{[s-d]}} - [\hat{\phi}^{\text{[s-d]}}]^{[k]}\big) \frac{j2 \pi m}{N} e^{\frac{j2\pi m [\hat{\phi}^{\text{[s-d]}}]^{[k]} }{N}},
\end{split}
\end{equation}
where $[\hat{\phi}^{\text{[s-d]}}]^{[k]}$ is the estimated CFO at the $k$-th iteration. Using \eqref{III-19}, $\mathcal{L}_{{\phi}^{\text{[s-d]}}}$ in \eqref{III-18-1} can be approximated as\vspace{-0pt}
\begin{align}\label{III-20}
\mathcal{L}_{{\phi}^{\text{[s-d]}}}  \approx&  \Big({\bf y}^{\text{[s]}}-[\hat{{\bf \Lambda}}^{[k]}_{\phi^{\text{[s-d]}}} + (\phi^{\text{[s-d]}} - [\hat{\phi}^{\text{[s-d]}}]^{[k]})\tilde{{\bf \Lambda}}^{[k]}_{\phi^{\text{[s-d]}}} ]{\bf d}^{\text{[s-d]}} \Big)^H \notag\\
&\times 
\Big[[\hat{{\bf \Sigma}}^{\text{[r]}}]^{[k]} \Big]^{-1}
\Big({\bf y}^{\text{[s]}}-[\hat{{\bf \Lambda}}^{[k]}_{\phi^{\text{[s-d]}}} \notag\\
&\;\;\;\;\;\;\;\;\;\;\;\;\;+ \big(\phi^{\text{[s-d]}} - [\hat{\phi}^{\text{[s-d]}}]^{[k]}\big)\tilde{{\bf \Lambda}}^{[k]}_{\phi^{\text{[s-d]}}} ]{\bf d}^{\text{[s-d]}} \Big),
\end{align}
where ${\bf d}^{\text{[s-d]}}\triangleq  \alpha \hat{{\bm \Lambda}}^{[k+1]}_{{\bm\theta}^{\text{[s-d]}}}{\bf F}^H {\bm \Lambda}_{s^{\text{[s]}}} {\bf F}_{[L]} \hat{{\bf c}}^{[k+1]}$ and $\tilde{{\bf \Lambda}}^{[k]}_{\phi^{\text{[s-d]}}}$ is a diagonal matrix where its $m$-th diagonal element is given by $[\tilde{{\bf \Lambda}}_{\phi^{\text{[s-d]}}}]_{m,m}=\frac{j2\pi (m-1)}{N} e^{\frac{j2\pi [\hat{\phi}^{\text{[s-d]}}]^{[k]} (m-1)}{N}}$. By setting $\frac{\partial \mathcal{L}_{{\phi}^{\text{[s-d]}}}}{\partial \phi^{\text{[s-d]}}}=0$ and solving for $\phi^{\text{[s-d]}}$, a closed-form solution for the CFO estimate at the $(k+1)$-th iteration, $[\phi^{\text{[s-d]}}]^{k+1}$, can be found as
\begin{equation}\label{III-21}
\begin{split}
& [\hat{\phi}^{\text{[s-d]}}]^{k+1} = [\hat{\phi}^{\text{[s-d]}}]^{k} + \\
& \frac{\Re\big( ({\bf y}^{\text{[s]}} - \hat{{\bf \Lambda}}^{[k]}_{\phi^{\text{[s-d]}}}{\bf d}^{\text{[s-d]}})^H
\left[[\hat{{\bf \Sigma}}^{\text{[r]}}]^{[k]} \right]^{-1}
\tilde{{\bf \Lambda}}^{[k]}_{\phi^{\text{[s-d]}}}{\bf d}^{\text{[s-d]}}\big)}{[{\bf d}^{\text{[s-d]}}]^H [\tilde{{\bf \Lambda}}^{[k]}_{\phi^{\text{[s-d]}}}]^H
\left[[\hat{{\bf \Sigma}}^{\text{[r]}}]^{[k]} \right]^{-1}
\tilde{{\bf \Lambda}}^{[k]}_{\phi^{\text{[s-d]}}} {\bf d}^{\text{[s-d]}}}.
\end{split}
\end{equation}
Finally, the noise covariance matrices ${\bf \Sigma}^{\text{[r]}}$, and $\hat{{\bf \Sigma}}^{[k]}$ are updated using $[\phi^{\text{[s-d]}}]^{k+1}$ as $[\hat{{\bf \Sigma}}^{\text{[r]}}]^{[k+1]}$ and $\hat{{\bf \Sigma}}^{[k+1]}$.

\noindent The overall iterative joint estimation algorithm can be summarized as follows:

\vspace{-0.2cm}
\hrulefill
\par
{\footnotesize
\textbf{Algorithm 1}
\begin{itemize}
\item \textbf{Solve} $\hat{\phi}^{\text{[r-d]}}$ and $\hat{{\bm \theta}}^{\text{[r-d]}}$ using \eqref{III-2} and \eqref{III-3} and
initialize $\phi^{\text{[s-d]}}$, ${\bf g}$, ${\bf h}$, ${\bf \Sigma}^{\text{[r]}}$.
\item \textbf{Repeat}
\begin{itemize}
\item Update $[\hat{{\bm \theta}}^{\text{[s-d]}}]^{[k+1]}$ with $[{\hat{\phi}^{\text{[s-d]}}}]^{[k]}$, $\hat{{\bf h}}^{[k]}$ and $\hat{{\bf g}}^{[k]}$ being fixed by using \eqref{III-10} and then update $[\hat{{\bf \Sigma}}^{\text{[r]}}]^{[k]}$;
\item Update $\hat{{\bf g}}^{[k+1]}$ with $[\hat{{\phi}}^{\text{[s-d]}}]^{[k]}$, $\hat{{\bf h}}^{[k]}$ and $[\hat{{\bm \theta}}^{\text{[s-d]}}]^{[k+1]}$ being fixed by using \eqref{III-16} and then update $[\hat{{\bf \Sigma}}^{\text{[r]}}]^{[k]}$;
\item Update $\hat{{\bf h}}^{[k+1]}$ with $[\hat{\phi}^{\text{[s-d]}}]^{[k]}$, $\hat{{\bf g}}^{[k+1]}$ and $[\hat{{\bm \theta}}^{\text{[s-d]}}]^{[k+1]}$ being fixed by using \eqref{III-18};
\item Update $[\hat{\phi}^{\text{[s-d]}}]^{[k+1]}$ with $\hat{{\bf h}}^{[k+1]}$, $\hat{{\bf g}}^{[k+1]}$ and $[\hat{{\bm \theta}}^{\text{[s-d]}}]^{[k+1]}$ being fixed by using \eqref{III-21} and then update $[\hat{{\bf \Sigma}}^{\text{[r]}}]^{[k]}$ as $[\hat{{\bf \Sigma}}^{\text{[r]}}]^{[k+1]}$;
\end{itemize}
\item \textbf{Until} $e(n+1)-e(n) \leq \varepsilon$ where $e(n)$ denotes the obtained value of objective function in \eqref{V-2} after the $n$-th iteration and $\varepsilon$ is a pre-set convergence accuracy.
\end{itemize}}
\vspace{-0.3cm}
\hrulefill

\vspace{-14pt}
\subsection{Initialization of the Proposed Iterative Algorithm}\label{sec:Initialization}
In \emph{Algorithm 1}, initial estimates of the CFO, relay-to-declination channel, source-to-relay channel, and $\hat{{\bf \Sigma}}^{\text{[r]}}$, which are denoted by $[\hat{\phi}^{\text{[s-d]}}]^{[0]}$, $\hat{\bf g}^{[0]}$, $\hat{\bf h}^{[0]}$ and $[\hat{{\bf \Sigma}}^{\text{[r]}}]^{[0]}$, respectively, are required. Thus, we present the initialization steps for the proposed iterative estimator. Simulations in Section \ref{sec:simulations} show that the proposed estimator converges to the true values of the parameters of interest for this choice of initialization. 

Since the relay-to-destination CFO and PN parameters, $\hat{\phi}^{\text{[r-d]}}$ and $\hat{\bm \theta}^{\text{[r-d]}}$, respectively, are estimated via \eqref{III-2} and \eqref{III-3}, respectively, the initial relay-to-destination channel estimates, $\hat{{\bf g}}^{[0]}$, can be obtained from the received signal ${\bf y}^{\text{[r]}}$ via $\hat{{\bf g}}^{[0]} = \frac{1}{{N}P^{\text{[r]}}_{\text{T}}} {\bf F}^H_{[L_g]} {\bm \Lambda}^H_{s^{\text{[r]}}}{\bf F} \hat{{\bm \Lambda}}^H_{\theta^{\text{[r-d]}}} \hat{{\bm \Lambda}}^H_{\phi^{\text{[r-d]}}} {\bf y}^{\text{[r]}}$ \cite{Darryl2006}. Next, we seek to obtain the initial estimates of the source-to-destination CFO, $[\hat{\phi}^{\text{[s-d]}}]^{[0]}$, and source-to-relay channel, $\hat{{\bf h}}^{[0]}$. By ignoring the PN terms, \eqref{III-1} can be approximated as \vspace{-4pt}
\begin{equation}\label{III-23} \nonumber
\begin{split}
{\bf y}^{\text{[s]}} \approx {\alpha} {\bf \Lambda}_{\phi^{\text{[s-d]}}}  {\bf F}^H {\bm \Lambda}_{s^{\text{[s]}}} {\bf F}_{[L]} \hat{\tilde{{\bf G}}}^{[0]} {\bf h} +  {\alpha} {\bf \Lambda}_{\phi^{\text{[s-d]}}}  \hat{{\bf G}}^{[0]} {\bf v}  + {\bf w},
\end{split}
\end{equation}
where $\hat{\tilde{{\bf G}}}^{[0]}$ and $\hat{{\bf G}}^{[0]}$ are formed via $\hat{{\bf g}}^{[0]}$ according to (\ref{III-12}) and (\ref{EQU-10-1}), respectively. Subsequently, using the ML criterion the initial estimates of the CFO, $[\hat{\phi}^{\text{[s-d]}}]^{[0]}$, and channel, $\hat{{\bf h}}^{[0]}$, can be obtained by minimizing \vspace{-4pt}
\begin{equation}\label{III-24} \nonumber
\begin{split}
  \{ \hat{{\bf h}}^{[0]}, [\hat{\phi}^{\text{[s-d]}}]^{[0]}  \} = & \min_{{\bf h}, {\phi^{\text{[s-d]}}}} ~({\bf y}^{\text{[s]}} - {\alpha} {\bf \Lambda}_{\phi^{\text{[s-d]}}}  {\bf F}^H {\bm \Lambda}_{s^{\text{[s]}}} {\bf F}_{[L]} \hat{\tilde{{\bf G}}}^{[0]} {\bf h} )^H \\
 & \times \left[{\bf \Sigma}^{\text{[r]}} \right]^{-1}
 ({\bf y}^{\text{[s]}} - {\alpha} {\bf \Lambda}_{\phi^{\text{[s-d]}}}  {\bf F}^H {\bm \Lambda}_{s^{\text{[s]}}} {\bf F}_{[L]} \hat{\tilde{{\bf G}}}^{[0]} {\bf h} )  \\
 & + \log \det({\bf \Sigma}^{\text{[r]}}),
\end{split}
\end{equation}
where ${\bf \Sigma}^{\text{[r]}} = {\alpha}^2 \sigma^2_R {\bf \Lambda}_{\phi^{\text{[s-d]}}} \hat{{\bf G}}^{[0]} [\hat{{\bf G}}^{[0]}]^H {\bf \Lambda}^H_{\phi^{\text{[s-d]}}}+ \sigma^2_D {\bf I}_N $. Accordingly, $[\hat{\phi}^{\text{[s-d]}}]^{[0]}$ and $\hat{{\bf h}}^{[0]}$ can be determined as \cite{StevenBook}
\begin{align}\label{III-25}
 \hspace{-6pt}\hat{{\bf h}}^{[0]} = & \left({\alpha} [\hat{\tilde{{\bf G}}}^{[0]}]^H {\bf F}^H_{[L]} {\bm \Lambda}^H_{s^{\text{[s]}}}{\bf F} {\bf \Lambda}^H_{\phi^{\text{[s-d]}}}
 \left[ {\bf \Sigma}^{\text{[r]}}  \right]^{-1}
 {\bf \Lambda}_{\phi^{\text{[s-d]}}}  {\bf F}^H {\bm \Lambda}_{s^{\text{[s]}}}  \right. \notag\\
& \left. \times  {\bf F}_{[L]} \hat{\tilde{{\bf G}}}^{[0]} \right)^{-1}
 [\hat{\tilde{{\bf G}}}^{[0]}]^H {\bf F}^H_{[L]} {\bm \Lambda}^H_{s^{\text{[s]}}}{\bf F} {\bf \Lambda}^H_{\phi^{\text{[s-d]}}}
 \left[ {\bf \Sigma}^{\text{[r]}}  \right]^{-1}
{\bf y}^{\text{[s]}}, \notag\\
 \hspace{-6pt}[\hat{\phi }^{\text{[s-d]}}]^{[0]}=  & \min_{{\phi^{\text{[s-d]}}}}~ ({\bf y}^{\text{[s]}} - {\alpha}{\bf \Lambda}_{\phi^{\text{[s-d]}}}  {\bf F}^H {\bm \Lambda}_{s^{\text{[s]}}} {\bf F}_{[L]} \hat{\tilde{{\bf G}}}^{[0]} \hat{{\bf h}}^{[0]} )^H \\
& \times \left[ {\bf \Sigma}^{\text{[r]}} \right]^{-1}
 ({\bf y}^{\text{[s]}} - {\alpha} {\bf \Lambda}_{\phi^{\text{[s-d]}}}  {\bf F}^H {\bm \Lambda}_{s^{\text{[s]}}} {\bf F}_{[L]} \hat{\tilde{{\bf G}}}^{[0]} \hat{{\bf h}}^{[0]} ),
 \notag
\end{align}
where the minimization in \eqref{III-25} is carried out through a one-dimensional exhaustive search. Although this process can be computationally intensive, it is only required to be carried out at the initial setup, since for subsequent OFDM packets, the previous CFO estimates can be applied to initialize the proposed iterative estimator. As for the additive noise covariance matrix, $[\hat{{\bf \Sigma}}^{\text{[r]}}]^{[0]}$, using the Taylor approximation in Section \ref{sec:PN}, we have\vspace{-0pt}
\begin{align}\label{III-26-1}
[\hat{{\bf \Sigma}}^{\text{[r]}}]^{[0]} &= {\alpha}^2 \sigma^2_R {\bf \Lambda}_{{\bm \theta}^{\text{[s-d]}}} \hat{{\bf \Lambda}}_{\phi^{\text{[s-d]}}} \hat{{\bf G}}^{[0]} [\hat{{\bf G}}^{[0]}]^H \hat{{\bf \Lambda}}^H_{\phi^{\text{[s-d]}}} {\bf \Lambda}^H_{{\bm \theta}^{\text{[s-d]}}}+ \sigma^2_D {\bf I}_N \notag\\
& \approx {\bf \Omega} +  {\bf \Omega}\odot \left( {\bm \theta}^{\text{[s-d]}} [{\bm \theta}^{\text{[s-d]}}]^H \right) + \sigma^2_D {\bf I}_N \\
& \approx {\bf \Omega} +  {\bf \Omega}\odot {\bf \Psi}^{\text{[s-d]}} + \sigma^2_D {\bf I}_N,\notag
\end{align}
where $ {\bf \Omega}\triangleq {\alpha}^2 \sigma^2_R \hat{{\bf \Lambda}}_{\phi^{\text{[s-d]}}} \hat{{\bf G}}^{[0]} [\hat{{\bf G}}^{[0]}]^H \hat{{\bf \Lambda}}^H_{\phi^{\text{[s-d]}}} $. In \eqref{III-26-1}, since ${\bm \theta}^{\text{[s-d]}}$ is not known, we use the expectation $\mathbb{E}({\bm \theta}^{\text{[s-d]}} [{\bm \theta}^{\text{[s-d]}}]^H) = {\bf \Psi}^{\text{[s-d]}}$ instead of the term ${\bm \theta}^{\text{[s-d]}} [{\bm \theta}^{\text{[s-d]}}]^H$. This allows for a closed-form expression for obtaining the source-to-relay channel estimates. 


\newtheorem{remark}{Remark}

\begin{remark}\label{rem:0}
Similar to point-to-point systems \cite{Darryl2007, Jun2009, Septier2008}, while jointly estimating the channel, CFO, and PN parameters in OFDM relay systems, a residual ambiguity may exist amongst these parameters. In what follows, we demonstrate the impact of this ambiguity on evaluating the performance of the proposed estimators.

The negative LLF in \eqref{III-6} can be rewritten as
\begin{align}\label{III-29}
\{\hat{\phi}^{\text{[s-d]}}, \hat{{\bm \theta}}^{\text{[s-d]}}, \hat{\phi}^{\text{[r-d]}}, \hat{{\bm \theta}}^{\text{[r-d]}}, \hat{{\bf h}}, \hat{{\bf g}}\}
\propto&  \arg \min \log \det({\bf \Sigma})\\
&+({\bf y} - {\bm \mu})^H {\bf \Sigma}^{-1} ({\bf y} - {\bm \mu}) \notag\\
&+ \frac{1}{2}[{\bm \theta}^{\text{[s-d]}}]^T [{\bf \Psi}^{\text{[s-d]}}]^{-1} {\bm \theta}^{\text{[s-d]}} \notag\\
&+ \frac{1}{2}[{\bm \theta}^{\text{[r-d]}}]^T [{\bf \Psi}^{\text{[r-d]}}]^{-1} {\bm \theta}^{\text{[r-d]}}.\notag
\end{align}
Eq. \eqref{III-29} is similar to \eqref{III-6} with the exception that ${\phi^{\text{[r-d]}}}$ and ${\bm \theta}^{\text{[r-d]}}$ are also treated as parameters of interest and ${\bm \eta}^{[i]}$ is replaced with ${\bm \theta}^{[i]}$. At very high SNR, i.e., $\sigma^2_R \rightarrow 0$ and $\sigma^2_D \rightarrow 0$, \eqref{III-29} can be further simplified as\vspace{-0pt}
\begin{align}\label{III-30}
\hspace{-5pt}\{\hat{\phi}^{\text{[s-d]}}, \hat{{\bm \theta}}^{\text{[s-d]}}, \hat{\phi}^{\text{[r-d]}}, \hat{{\bm \theta}}^{\text{[sr-d]}}, \hat{{\bf h}}, \hat{{\bf g}}\} \propto& \arg \min
  \log \det({\bf \Sigma})\\
  &+({\bf y} - {\bm \mu})^H {\bf \Sigma}^{-1} ({\bf y} - {\bm \mu}).\notag
\end{align}
From \eqref{III-30} it can be concluded that the metric for estimation of parameters of interest is solely dependent on the received signal instead of the prior information at high SNR \cite{StevenBook}. Moreover, it can be straightforwardly shown that the received training symbols, e.g., ${\bf y}^{\text{[r]}}$, are not altered under a common phase rotation, $\varphi_g$, between the channel response, $\hat{{\bf g}}$, and PN parameters $\hat{\bm \theta}^{\text{[r-d]}}$, i.e.,\vspace{-6pt}
\begin{equation}\label{III-27}
\begin{split}
\hat{{\bf g}} \rightarrow \exp(-j\varphi_g){\bf g},~~
\hat{{\bm \theta}}^{\text{[r-d]}} \rightarrow {\bm \theta}^{\text{[r-d]}} + \varphi_g {\bf 1}_N.
\end{split}
\end{equation}
Thus, the common phase rotation, $\varphi_g$, can be considered as a phase ambiguity amongst the channel and PN parameters that cannot be estimated. Using a similar approach, it can also be shown that there exists a phase ambiguity between the estimate of the source-to-relay channel, $\hat{{\bf h}}$, and the estimate of the source-to-destination PN parameter, $\hat{\bm \theta}^{\text{[s-d]}}$ given by\vspace{-0pt}
\begin{equation}\label{III-27-1}
\begin{split}
\hat{{\bf h}} \rightarrow \exp(-j\varphi_h){\bf h},~~
\hat{{\bm \theta}}^{\text{[s-d]}} \rightarrow {\bm \theta}^{\text{[s-d]}} + (\varphi_h+\varphi_g) {\bf 1}_N,
\end{split}
\end{equation}
where $\varphi_h$ is the phase ambiguity associated with channel ${{\bf h}}$. 
In addition to the ambiguity between channel and PN, a phase ambiguity may also exist between the PN and CFO as: 
\begin{equation}\label{III-28}
\begin{split}
\hat{\phi}^{\text{[s-d]}} &\rightarrow \phi^{\text{[s-d]}} - \epsilon^{\text{[s-d]}},\;\;\;\;\;\;
\hat{\phi}^{\text{[r-d]}} \rightarrow \phi^{\text{[r-d]}} - \epsilon^{\text{[r-d]}},\\
\hat{{\bm \theta}}^{\text{[s-d]}} &\rightarrow {\bm \theta}^{\text{[s-d]}} + {\bm \epsilon}^{\text{[s-d]}},\;\;\;\;\;\;
\hat{{\bm \theta}}^{\text{[r-d]}} \rightarrow {\bm \theta}^{\text{[r-d]}} + {\bm \epsilon}^{\text{[r-d]}},\\
\end{split}
\end{equation}
where $[{\bm \epsilon}^{\text{[s-d]}}]_m = \frac{2\pi (m-1)\epsilon^{\text{[s-d]}}}{N}$ and $[{\bm \epsilon}^{\text{[r-d]}}]_m = \frac{2\pi (m-1)\epsilon^{\text{[r-d]}}}{N}$.
These ambiguities make it difficult to assess the estimation accuracy of the proposed iterative estimator. Thus, here, a new approach for determining the MSE of the estimated parameters is proposed. The MSE of the channel responses $\hat{\bf h}$ and $\hat{\bf g}$, can be computed as
\begin{equation}\label{III-31}
\begin{split}
{\rm MSE}_{\bf g} = ||\underline{\hat{\bf g}} - \underline{{\bf g} } ||^2_2,~~{\rm MSE}_{\bf h} = ||\underline{\hat{\bf h}} - \underline{{\bf h} } ||^2_2,
\end{split}
\end{equation}
where $\underline{\hat{\bf g}} \triangleq \exp(-j \angle \hat{ g}(0))\hat{\bf g}$,  $\underline{\hat{\bf h}} \triangleq \exp( -j \angle \hat{ h}(0))\hat{\bf h}$, $\underline{{\bf g}} \triangleq \exp(-j \angle { g}(0)){\bf g}$ and $\underline{{\bf h}} \triangleq \exp( -j \angle { h}(0)){\bf h}$.
Using this approach, the phase ambiguity between the PN and channels, does not affect the MSE of channel estimation. Similarly, for the CFO and PN, the overall MSE is calculated as
\begin{equation}\label{III-32}
\begin{split}
{\rm MSE}_{\phi^{\text{[s-d]}}, {\bm \theta}^{\text{[s-d]}}} = ||\underline{{\bm \delta}}- \underline{\hat{\bm \delta}} ||^2_2,
\end{split}
\end{equation}
where $\underline{{\bm \delta}} = {\bm \delta} - {\delta}_0 {\bf 1}$, $\underline{\hat{\bm \delta}} = \hat{\bm \delta} - \hat{\delta}_0 {\bf 1}$, ${\bm \delta} =[\delta_0,\delta_0,\cdots,\delta_{N-1}]^T$ with $\delta_m = {\theta}^{\text{[s-d]}}(m) + \frac{2\pi (m-1)\phi^{\text{[s-d]}}}{N}$, and $\hat{\bm \delta}=[\hat{\delta}_0,\hat{\delta}_1,\cdots,\hat{\delta}_{N-1}]^T$ with $\hat{\delta}_m = \hat{\theta}^{\text{[s-d]}}(m) + \frac{2\pi (m-1){\hat\phi}^{\text{[s-d]}}}{N}$.

\end{remark}

\vspace{-4pt}
\section{The Hybrid Cram\'{e}r-Rao Lower Bound}\label{sec:HCRLB}
In this section, a the HCRLB for joint estimation of channel, CFO, and PN in OFDM relay networks is derived. 

As stated in \emph{Remark \ref{rem:0}}, due to the ambiguities between the estimation of channel responses, CFO, and PN, \eqref{III-1} and \eqref{III-1-1} are first rewritten as
\begin{equation}\label{III-33}
\begin{split}
{\bf y}^{\text{[s]}} &=  \alpha {\bm \Lambda}_{\bm\theta^{\text{[s-d]}}} {\bm \Lambda}_{\phi^{\text{[s-d]}}} \left( {\bf F}^H \underline{{\bm \Lambda}}_{s^{\text{[s]}}} {\bf F}_{[L]}\underline{{\bf c}} + \underline{{\bf G}} {\bf v} \right) + {\bf w}, \\
{\bf y}^{\text{[r-d]}} &={\bm \Lambda}_{\bm\theta^{\text{[r-d]}}} {\bm \Lambda}_{\phi^{\text{[r-d]}}}  {\bf F}^H \underline{{\bm \Lambda}}_{s^{\text{[r]}}} {\bf F}_{[L_g]}\underline{{\bf g}} + {\bf w},
\end{split}
\end{equation}
where $\underline{{\bf c}} \triangleq \underline{{\bf h}}\star \underline{{\bf g}}$ with $\underline{{\bf h}}$ and $ \underline{{\bf g}}$ defined in \eqref{III-31}, $[\underline{{\bm \Lambda}}_{s^{\text{[s]}}}]_{m,m} \triangleq{s}^{\text{[s]}}_{m-1} \exp\left(j\angle { c}(0)\right)$, $[\underline{{\bm \Lambda}}_{s^{\text{[r]}}}]_{m,m} \triangleq{s}^{\text{[r]}}_{m-1} \exp \left( j\angle { g}(0)\right)$ are known diagonal training signal matrices that are rotated by the phases of the first elements of the channels, ${\bf c}$ and ${\bf g}$, respectively, and matrix $\underline{{\bf G}}$ is constructed using $ \underline{{\bf g}}$ similar to (\ref{EQU-10-1}). Accordingly, the HCRLB for the estimation problem is given by \cite{Trees2001}\vspace{-0pt}
\begin{equation}\label{III-33-1}\nonumber
\begin{split}
\mathbb{E}_{{\bf y}, {\bm \theta}^{\text{[s-d]}}, {\bm \theta}^{\text{[r-d]}} |\phi^{\text{[s-d]}},\phi^{\text{[r-d]}},\underline{{\bf g}},\underline{{\bf h}} } \left[({\bm \lambda} - \hat{\bm \lambda})({\bm \lambda} - \hat{\bm \lambda})^T   \right] \succeq {\bf B}^{-1},
\end{split}
\end{equation}
where ${\bm \lambda}\triangleq\left[\phi^{\text{[s-d]}},({\bm \theta}^{\text{[s-d]}})^T, \phi^{\text{[r-d]}}, ({\bm \theta}^{\text{[r-d]}})^T, \underline{g}_0, \Re(\underline{\tilde{{\bf g}}})^T, \Im(\underline{\tilde{{\bf g}}})^T,\right.$ $\left. \underline{h}_0, \Re(\underline{\tilde{{\bf h}}})^T, \Im(\underline{\tilde{{\bf h}}})^T \right]^T$ denotes the vector of parameters of interest, $\underline{\tilde{{\bf g}}} \triangleq \underline{\bf g}(1:L_g-1)$, $\underline{\tilde{{\bf h}}} \triangleq \underline{\bf h}(1:L_h-1)$, and ${\bf B}$ is the Bayesian information matrix (BIM) that is given by
\begin{align}\label{III-33-2}
{\bf B} =& \mathbb{E}_{{\bm \theta}^{\text{[s-d]}}, {\bm \theta}^{\text{[r-d]}}}\left[ {\bf FIM}({\bf y};{\bm \lambda}) \right] +
\mathbb{E}_{{\bm \theta}^{\text{[s-d]}}, {\bm \theta}^{\text{[r-d]}}} \left[- \triangle^{{\bm \lambda}}_{{\bm \lambda}}\log p({\bm \theta}^{\text{[s-d]}})  \right] \notag \\
&+ \mathbb{E}_{{\bm \theta}^{\text{[s-d]}}, {\bm \theta}^{\text{[r-d]}}} \left[- \triangle^{{\bm \lambda}}_{{\bm \lambda}} \log p({\bm \theta}^{\text{[r-d]}})  \right].
\end{align}
In \eqref{III-33-2}, ${\bf FIM}({\bf y};{\bm \lambda}) = \mathbb{E}_{{\bf y}}\left[ - \triangle^{{\bm \lambda}}_{{\bm \lambda}} \log p({\bf y};{\bm \lambda}) \right]$ denotes the Fisher's information matrix (FIM). In the following subsection the BIM in \eqref{III-33-2} is derived in detail.

\vspace{-4pt}
\subsection{Derivation of $\mathbb{E}_{{\bm \theta}^{\text{[s-d]}}, {\bm \theta}^{\text{[r-d]}}}\left[ {\bf FIM}({\bf y};{\bm \lambda}) \right] $}\label{sec:FIMfirst_term}
In order to derive $\mathbb{E}_{{\bm \theta}^{\text{[s-d]}}, {\bm \theta}^{\text{[r-d]}}}\left[ {\bf FIM}({\bf y};{\bm \lambda}) \right] $, we first derive the FIM for the parameters of interest ${\bm\lambda}$.


\textbf{Theorem} 1: The $Q \times Q$ Fisher's information matrix ${\bf FIM}({\bf y};{\bm \lambda})$ with $Q = 2(N+L_g+L_h)$ for the joint estimation problem is given by
\begin{equation}\label{III-Add-1}
\begin{split}
{\bf FIM}({\bf y};{\bm \lambda}) = \left[
  \begin{array}{ccc}
    {\rm FIM}_{1,1} + {\Upsilon}_{1,1} & \cdots & {\rm FIM}_{1,Q} + {\Upsilon}_{1,Q} \\
    \vdots & \ddots & \vdots \\
    {\rm FIM}_{Q,1} + {\Upsilon}_{Q,1} & \cdots & {\rm FIM}_{Q,Q} + {\Upsilon}_{Q,Q} \\
  \end{array}
 \right].
\end{split}
\end{equation}
In \eqref{III-Add-1}, ${\rm FIM}_{i,j}$, for $i,j=1,2,\cdots,Q$, is determined as
\begin{equation}\label{III-Add-2}
\begin{split}
{\rm FIM}_{i,j} = 2 \Re \left( {\bm \rho}^H_i \underline{{\bf \Sigma}}^{-1} {\bm \rho}_j  \right),
\end{split}
\end{equation}
where $\underline{{\bf \Sigma}} = {\rm Blkdiag}(\underline{{\bf \Sigma}}^{\text{[r]}}, \sigma^2_D {\bf I}_N)$ with $\underline{{\bf \Sigma}}^{\text{[r]}}=\alpha^2 \sigma^2_R {\bm \Lambda}_{\bm\theta^{\text{[s-d]}}} $\\${\bm \Lambda}_{\phi^{\text{[s-d]}}} \underline{{\bf G}}\underline{{\bf G}}^H  {\bm \Lambda}^H_{\phi^{\text{[s-d]}}} {\bm \Lambda}^H_{\bm\theta^{\text{[s-d]}}} + \sigma^2_D {\bf I}_N$, and ${\bm \rho}_i $ is given by
\begin{itemize}
\item $i=1$\vspace{-6pt}
\begin{equation}\nonumber
\begin{split}
\hspace{-30pt}{\bm \rho}_i \triangleq \left[(\alpha{\bm \Lambda} {\bm \Lambda}_{\bm\theta^{\text{[s-d]}}} {\bm \Lambda}_{\phi^{\text{[s-d]}}}  {\bf F}^H \underline{{\bm \Lambda}}_{s^{\text{[s]}}} {\bf F}_{[L]}\underline{{\bf c}})^T,
                                                             {\bf 0}^T_{N \times 1}
                                                         \right]^T,
\end{split}
\end{equation}
where ${\bm \Lambda}$ is a diagonal matrix with $[{\bm \Lambda}]_{m,m} = \frac{j2\pi (m-1)}{N}$;
\item $i=2,3,\cdots, N+1$\vspace{-6pt}
\begin{equation}\nonumber
\begin{split}
\hspace{-10pt}{\bm \rho}_i \triangleq \left[
                                                             ({\rm Diag} \left(\alpha {\bm \Lambda}_{\bm\phi^{\text{[s-d]}}}  {\bf F}^H \underline{{\bm \Lambda}}_{s^{\text{[s]}}} {\bf F}_{[L]}\underline{{\bf c}}\right){\mathbf a}_{i-2} )^T,
                                                             {\bf 0}^T_{N \times 1}
                                                         \right]^T,
\end{split}
\end{equation}
where ${\bf a}_{m}\triangleq[{\bf 0}_{1,m-1}, j\exp(j\theta^{\text{[s-d]}}(m)), {\bf 0}_{1,N-m}]^T$;
\item $i=N+2$\vspace{-6pt}
\begin{equation}\nonumber
\begin{split}
\hspace{-35pt}{\bm \rho}_{i} \triangleq  \left[
                                                             {\bf 0}^T_{N \times 1},
                                                             ({\bm \Lambda} {\bm \Lambda}_{\bm\theta^{\text{[r-d]}}} {\bm \Lambda}_{\phi^{\text{[r-d]}}}  {\bf F}^H \underline{{\bm \Lambda}}_{s^{\text{[r]}}} {\bf F}_{[L_g]}\underline{{\bf g}})^T                                                          \right]^T;
\end{split}
\end{equation}
\item $i=N+3,N+3,\cdots,2N+2$\vspace{-6pt}
\begin{equation}\nonumber
\begin{split}
{\bm \rho}_i \triangleq \left[
                                                             {\bf 0}^T_{N \times 1},
                                                             ({\rm Diag} \left({\bm \Lambda}_{\phi^{\text{[r-d]}}}  {\bf F}^H \underline{{\bm \Lambda}}_{s^{\text{[r]}}} {\bf F}_{[L_g]}\underline{{\bf g}}\right){\bf b}_{i-N-3})^T
                                                         \right]^T,
\end{split}
\end{equation}
where ${\bf b}_{m}\triangleq[{\bf 0}_{1,m-1}, j\exp(j\theta^{\text{[r-d]}}(m)), {\bf 0}_{1,N-m}]^T$;
\vspace{+2pt}
\item $i=2N+3$\vspace{-6pt}
\begin{equation}\nonumber
\begin{split}
\hspace{-115pt}{\bm \rho}_i \triangleq {\bf E}(:,1);
\end{split}
\end{equation}
\item $i=2N+4, 2N+5,\cdots, 2N+L_h+2$\vspace{-6pt}
\begin{equation}\nonumber
\begin{split}
\hspace{-115pt}{\bm \rho}_i \triangleq{\bf E}(:,i-2N-2);
\end{split}
\end{equation}
\item $i=2N+L_h+3, 2N+L_h+4,\cdots, 2N+2L_h+1$\vspace{-6pt}
\begin{equation}\nonumber
\begin{split}
\hspace{-115pt}{\bm \rho}_i \triangleq j{\bf E}(:,i-2N-L_h-1),
\end{split}
\end{equation}
where
\begin{equation}\label{III-Add-2-1}
\begin{split}
{\bf E} \triangleq \left[
                            \begin{array}{c}
                              \alpha {\bm \Lambda}_{\phi^{\text{[s-d]}}} {\bm \Lambda}_{{\bm \theta}^{\text{[s-d]}}} {\bf F}^H \underline{{\bm \Lambda}}_{s^{\text{[s]}}} {\bf F}_{[L]}\underline{\tilde{{\bf H}}} \\
                              {\bm \Lambda}_{\phi^{\text{[r-d]}}} {\bm \Lambda}_{{\bm \theta}^{\text{[r-d]}}} {\bf F}^H \underline{{\bm \Lambda}}_{s^{\text{[r]}}} {\bf F}_{[L_g]} \\
                            \end{array}
\right]
%
\end{split}
\end{equation}
and $\underline{\tilde{{\bf H}}}$ is constructed via $\underline{{\bf h}}$ as in \eqref{III-12};\vspace{+2pt}
\item $i=2N+2L_h+2$\vspace{-6pt}
\begin{equation}\label{III-Add-2-1}
\begin{split}
\hspace{-115pt}{\bm \rho}_i \triangleq {\bf K}(:,1);
\end{split}
\end{equation}
\item $i=2N+2L_h+3,\cdots, 2N+2L_h+L_g+1$\vspace{-6pt}
\begin{equation}\label{III-Add-2-1}
\begin{split}
\hspace{-115pt}{\bm \rho}_i \triangleq {\bf K}(:,i-2N-2L_h-1);
\end{split}
\end{equation}
\item $i=2N+2L_h+L_g+2, \cdots, Q$\vspace{-6pt}
\begin{equation}\label{III-Add-2-1}
\begin{split}
\hspace{-105pt}{\bm \rho}_i \triangleq j{\bf K}(:,i-2N-2L_h-2L_g);
\end{split}
\end{equation}
\vspace{-18pt}\\
where\vspace{-6pt}
\begin{equation}\label{III-Add-2_5}
\begin{split}
\vspace{-6pt}
{\bf K} \triangleq \big[
                                                            (\alpha {\bm \Lambda}_{\phi^{\text{[s-d]}}} {\bm \Lambda}_{{\bm \theta}^{\text{[s-d]}}} {\bf F}^H \underline{{\bm \Lambda}}_{s^{\text{[s]}}} {\bf F}_{[L]}\underline{\tilde{{\bf G}}} )^T,
                                                             {\bf 0}^T_{N \times 1}
                                                         \big]^T,
\end{split}
\end{equation}
and $\underline{\tilde{{\bf G}}}$ is constructed using $\underline{{\bf g}}$ as in \eqref{III-11}.
\end{itemize}
\vspace{+4pt}
Moreover, in \eqref{III-Add-1}, for $i=1,2,\cdots,N+1$ and $j =2N+3,2N+4,\cdots,2N+2L_g+1$, ${\Upsilon}_{i,j} = {\rm Tr}\left[ \underline{{\bf \Sigma}}^{-1} {\bf Q}_i \underline{{\bf \Sigma}}^{-1} {\bf Q}_j \right]$ and ${\Upsilon}_{i,j} =0$ for all other $i$ and $j$. Note that ${\bf Q}_i \triangleq{\rm Blkdiag} \left( {\bf W}_i, {\bf 0}_{N\times N} \right) $ is given by
\begin{itemize}
  \item $i = 1$,\vspace{-4pt}
\begin{equation}\nonumber
\begin{split}
{\bf W}_i\triangleq  \big(\alpha^2 \sigma^2_R {\bm \Lambda}_{\bm\theta^{\text{[s-d]}}} \underline{{\bf G}}\underline{{\bf G}}^H  {\bm \Lambda}^H_{\bm\theta^{\text{[s-d]}}} \big)\odot
 \big({\bf \Lambda}{\bm \vartheta} {\bm \vartheta}^H + {\bm \vartheta} ({\bf \Lambda}{\bm \vartheta})^H \big);
\end{split}
\end{equation}
\vspace{-18pt}
 \item $i = 2,\cdots,N+1$\vspace{-2pt}
 \begin{equation}\nonumber
\begin{split}
\hspace{-75pt}{\bf W}_i \triangleq&  \big(\alpha^2 \sigma^2_R {\bm \Lambda}_{\bm\theta^{\text{[s-d]}}} \underline{{\bf G}}\underline{{\bf G}}^H  {\bm \Lambda}^H_{\bm\theta^{\text{[s-d]}}} \big)
\\
\hspace{-75pt}&\odot
 \big( {\bf a}_{i-2} [{\bm \theta}^{\text{[s-d]}}]^H +  {\bm \theta}^{\text{[s-d]}}{\bf a}^H_{i-2}  \big);
\end{split}
\end{equation}
\vspace{-8pt}
\item $i =2N+3$\vspace{-6pt}
 \begin{equation}\nonumber
\begin{split}
{\bf W}_i\triangleq \alpha^2 \sigma^2_R {\bm \Lambda}_{\bm\theta^{\text{[s-d]}}} {\bm \Lambda}_{\phi^{\text{[s-d]}}} \big({\bf D}_0 \underline{{\bf G}}^H + \underline{{\bf G}} {\bf D}^H_0 \big)  {\bm \Lambda}^H_{\phi^{\text{[s-d]}}} {\bm \Lambda}^H_{\bm\theta^{\text{[s-d]}}};
\end{split}
\end{equation}
\vspace{-16pt}
\item $i=2N+4,\cdots,2N+L_g+2$\vspace{-6pt}
\begin{equation}\nonumber
\begin{split}
\hspace{-10pt}{\bf W}_i\triangleq & \alpha^2 \sigma^2_R {\bm \Lambda}_{\bm\theta^{\text{[s-d]}}} {\bm \Lambda}_{\phi^{\text{[s-d]}}} \\
\hspace{-10pt}& \times \big( {\bf D}_{i-2N-3} \underline{{\bf G}}^H + \underline{{\bf G}} {\bf D}^H_{i-2N-3} \big)  {\bm \Lambda}^H_{\phi^{\text{[s-d]}}} {\bm \Lambda}^H_{\bm\theta^{\text{[s-d]}}};
\end{split}
\end{equation}
\vspace{-8pt}
\item $i=2N+L_g+3,\cdots,2N+2L_g+1$\vspace{-6pt}
\begin{equation}\nonumber
\begin{split}
\hspace{-15pt}{\bf W}_i\triangleq & j\alpha^2 \sigma^2_R {\bm \Lambda}_{\bm\theta^{\text{[s-d]}}} {\bm \Lambda}_{\phi^{\text{[s-d]}}} \\
\hspace{-15pt}& \times \big( {\bf D}_{i-L_g-2N-2} \underline{{\bf G}}^H - \underline{{\bf G}} {\bf D}^H_{i-2N-L_g-2}\big)  \\
\hspace{-15pt}&\times
{\bm \Lambda}^H_{\phi^{\text{[s-d]}}} {\bm \Lambda}^H_{\bm\theta^{\text{[s-d]}}};
\end{split}
\end{equation}
\end{itemize}
In the above, ${\bm \vartheta} \triangleq \big[1, e^{\frac{j2\pi \phi^{\text{[s-d]}}}{N}}, \cdots, e^{\frac{j2\pi (N-1)\phi^{\text{[s-d]}}}{N}} \big]$ and $
{\bf D}_m  \triangleq \left[{\bf 0}_{N \times (L_g-m-1)}, {\bf I}_N, {\bf 0}_{N \times m}\right]$, $\forall m$.

\begin{proof}
See Appendix~\ref{prof_Theorem1}.
\end{proof}

Although the FIM can be obtained in closed-form, a closed-form expression for $\mathbb{E}_{{\bm \theta}^{\text{[s-d]}}, {\bm \theta}^{\text{[r-d]}}}\left[ {\bf FIM}({\bf y};{\bm \lambda}) \right] $ cannot be obtained due to the presence of a complex multidimensional integration. Hence, here, $\mathbb{E}_{{\bm \theta}^{\text{[s-d]}}, {\bm \theta}^{\text{[r-d]}}}\left[ {\bf FIM}({\bf y};{\bm \lambda}) \right] $ is numerically evaluated.

\vspace{-7pt}
\subsection{Derivation of $\mathbb{E}_{{\bm \theta}^{\text{[s-d]}}, {\bm \theta}^{\text{[r-d]}}} \left[- \triangle^{{\bm \lambda}}_{{\bm \lambda}}\log p({\bm \theta}^{\text{[s-d]}})  \right]$ and $\mathbb{E}_{{\bm \theta}^{\text{[s-d]}}, {\bm \theta}^{\text{[r-d]}}} \left[- \triangle^{{\bm \lambda}}_{{\bm \lambda}} \log p({\bm \theta}^{\text{[r-d]}})  \right]$ }\label{sec:FIMsecond_term}
Since $p\left({\bm \theta}^{\text{[s-d]}}\right)$ and $p\left({\bm \theta}^{\text{[r-d]}}\right)$ are independent of $\phi^{\text{[s-d]}}$, $\phi^{\text{[r-d]}}$, $\underline{{\bf g}}$, and $\underline{{\bf h}}$, we can straightforwardly obtain
\begin{equation}\label{III-40}\nonumber
\begin{split}
& \mathbb{E}_{{\bm \theta}^{\text{[s-d]}}, {\bm \theta}^{\text{[r-d]}}} \left[- \triangle^{{\bm \lambda}}_{{\bm \lambda}}\log p({\bm \theta}^{\text{[s-d]}})  \right] =
{\rm Blkdiag}\Big(0,[{\bf \Psi}^{\text{[s-d]}}]^{-1},0, \\
& {\bf 0}_{N\times N},{\bf 0}_{(2L_g-1) \times (2L_g-1)}, {\bf 0}_{(2L_h-1) \times (2L_h-1)}\Big),\\
& \mathbb{E}_{{\bm \theta}^{\text{[s-d]}}, {\bm \theta}^{\text{[r-d]}}} \left[- \triangle^{{\bm \lambda}}_{{\bm \lambda}}\log p({\bm \theta}^{\text{[r-d]}})  \right] = {\rm Blkdiag}\Big(0,{\bf 0}_{N\times N},0,\\
& [{\bf \Psi}^{\text{[r-d]}}]^{-1},{\bf 0}_{(2L_g-1) \times (2L_g-1)}, {\bf 0}_{(2L_h-1) \times (2L_h-1)}\Big).
\end{split}
\end{equation}
Finally, the BIM in \eqref{III-33-2} can be calculated using the results in Sections \ref{sec:FIMfirst_term} and \ref{sec:FIMsecond_term}.

\vspace{-5pt}
\subsection{Derivation of the Transformed HCRLB}
As shown in \emph{Remark \ref{rem:0}}, due to the ambiguities in the estimation of parameters of interest, the MSE of the CFO and PN is computed jointly as shown in \eqref{III-32}. Consequently, the parameters of interests, ${\bm \lambda}$ need to be transformed to ${\bm \lambda}_{\text{mod}}=\left[\underline{{\bm \delta}}^T, \phi^{\text{[r-d]}}, ({\bm \theta}^{\text{[r-d]}})^T, \underline{g}_0, \Re(\underline{\tilde{{\bf g}}})^T, \Im(\underline{\tilde{{\bf g}}})^T,\underline{h}_0, \Re(\underline{\tilde{{\bf h}}})^T, \Im(\underline{\tilde{{\bf h}}})^T \right]^T$. Since $\delta_m = {\theta}^{\text{[s-d]}}(m) + \frac{2\pi (m-1)\phi^{\text{[s-d]}}}{N}$, this transformation can be written in matrix form as
\begin{align*}
{\bm \lambda}_{\text{mod}} ={\bf \Xi} {\bm \lambda},
\end{align*}
where ${\bf \Xi} \triangleq {\bf \Xi}_2 {\bf \Xi}_1 $, ${\bf \Xi}_1 \triangleq{\rm Blkdiag}\big(0,\tilde{{\bf \Xi}}_1,1,{\bf I}_{N\times N},$ ${\bf I}_{(2L_g-1) \times (2L_g-1)},{\bf I}_{(2L_h-1) \times (2L_h-1)}\big)$, ${\bf \Xi}_2 \triangleq{\rm Blkdiag}(\tilde{{\bf \Xi}}_2,1,$ ${\bf I}_{N\times N},{\bf I}_{(2L_g-1) \times (2L_g-1)}, {\bf I}_{(2L_h-1) \times (2L_h-1)})$, and
\begin{equation*}
\begin{split}
\tilde{{\bf \Xi}}_1 & \triangleq \left[
                        \begin{array}{ccccc}
                          0 & 0 & 0 & \cdots & 0 \\
                          -1 & 1 & 0 & \cdots & 0 \\
                          \vdots & \vdots & \vdots & \ddots & \vdots \\
                          -1 & 0 & 0 & 0 & 1 \\
                        \end{array}
                      \right] \in \mathbb{R}^{N \times N}, \\
\tilde{{\bf \Xi}}_2 & \triangleq \left[
                        \begin{array}{cccccc}
                          0 & 1 & 0 & 0 & \cdots & 0 \\
                          \frac{2\pi}{N} & 0 & 1  & 0 & \cdots & 0 \\
                          \vdots & \vdots & \vdots & \vdots & \ddots & \vdots \\
                           \frac{2\pi(N-1)}{N} & 0 & 0  & 0 & \cdots & 1 \\
                        \end{array}
                      \right] \in \mathbb{R}^{N \times (N+1)}.
\end{split}
\end{equation*}
Thus, the HCRLB for the transformed parameters of interest, ${\bm \lambda}_{\text{mod}}$, $\mathbf{HCRLB}_{\rm mod}$, is given by
$\mathbf{HCRLB}_{\rm mod} = {\bf \Xi} {\bf B}^{-1} {\bf \Xi}^T$ \cite{StevenBook}.

\vspace{-5pt}
\section{Data Detection in Presence of Phase Noise}

\begin{figure}[t]
\begin{centering}
\includegraphics[scale=0.52]{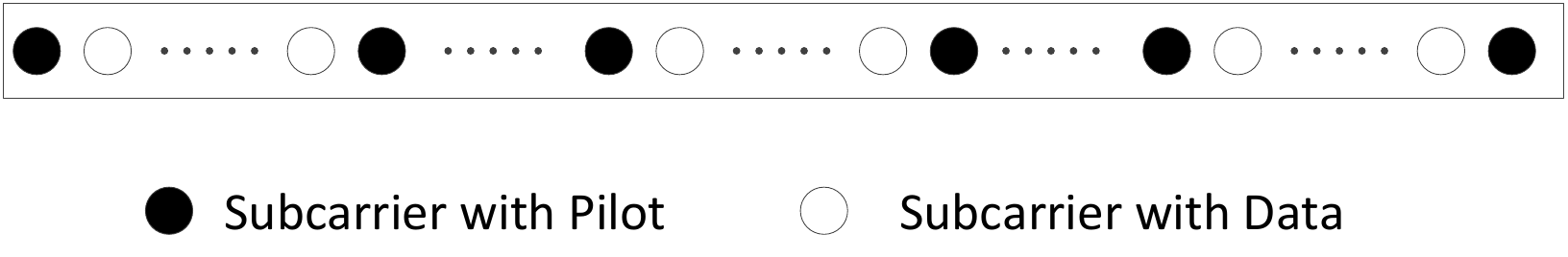}
\vspace{-0.1cm}
\caption{Illustration of the comb-type data symbol.} \label{Data_symbol}
\end{centering}
\vspace{-0.4cm}
\end{figure}

In this section, a receiver structure for data detection at the destination in the presence of PN is proposed. Since the PN parameters vary over an OFDM symbol, they need to be accurately tracked over the length of each symbol. Hence, we propose the transmission of comb-type data symbols from the source node, i.e., each transmitted symbol consists of both pilot and data subcarriers (see Fig.~\ref{Data_symbol}). As discussed in Section \ref{sec:short-PN}, in each OFDM data symbol, it is sufficient to estimate the shortened PN vector of length $M$. Thus, the number of subcarriers utilized for data transmission should be less than $N-M$. The received signal at the destination during the data transmission interval is given by
\begin{equation}\label{V-1}
\begin{split}
{\bf y}^{\text{[s]}} &= \alpha {\bm \Lambda}_{\bm\theta^{\text{[s-d]}}} {\bm \Lambda}_{\phi^{\text{[s-d]}}}  \left({\bf F} {\bf \Lambda}_{\tilde{\bf c}} {\bf s}^{\text{[s]}} + {\bf G} {\bf v} \right) + {\bf w} \\
 &= {\bf T}_{\text{T}} {\bf s}^{\text{[s]}}_{\text{T}} + {\bf T}_{\text{D}} {\bf s}^{\text{[s]}}_{\text{D}} + \alpha{\bm \Lambda}_{\bm\theta^{\text{[s-d]}}} {\bm \Lambda}_{\phi^{\text{[s-d]}}} {\bf G} {\bf v} + {\bf w},
\end{split}
\end{equation}
where ${\bf s}^{\text{[s]}}$ denotes the comb-type signal transmitted during the data transmission interval with $\mathbb{E} ({\bf s}^{\text{[s]}} \left[{\bf s}^{\text{[s]}}\right]^H ) = P^{\text{[s]}}_{\text{T}}{\bf I}_N$, ${\bf s}^{\text{[s]}}_{\text{T}}$ and ${\bf s}^{\text{[s]}}_{\text{D}}$ denote the pilot and data vector contained in ${\bf s}^{\text{[s]}}$, respectively, and ${\bf T}_{\text{T}}$ and ${\bf T}_{\text{D}}$ are the associated sub-matrices of the combined channel, $\alpha {\bm \Lambda}_{\bm\theta^{\text{[s-d]}}} {\bm \Lambda}_{\bm\phi^{\text{[s-d]}}}  {\bf F} {\bf \Lambda}_{\tilde{\bf c}}$, corresponding to ${\bf s}^{\text{[s]}}_{\text{T}}$ and ${\bf s}^{\text{[s]}}_{\text{D}}$, respectively. Since in \eqref{V-1}, the unknown PN vector ${\bm \theta}^{\text{[s-d]}}$ and data vector ${\bf s}^{\text{[s]}}_{\text{D}}$ are coupled with each other, similar to the estimation part, an iterative method is applied here. By using the MAP criterion as in \eqref{III-4}, the joint estimation of PN parameters and data can be formulated as
\begin{align}\label{V-2}
\{\hat{\bm \theta}^{\text{[s-d]}}, {\bf s}^{\text{[s]}}_{\text{D}}\} =& \arg \min_{{\bf \theta}^{\text{[s-d]}},  {\bf s}^{\text{[s]}}_{\text{D}}} \log \det({{\bf \Sigma}}^{\text{[r]}}) + \left({\bf y}^{\text{[s]}} - {\bm \mu}\right)^H \\
& \times \left[{{\bf \Sigma}}^{\text{[r]}}\right]^{-1} \left({\bf y}^{\text{[s]}} - {\bm \mu}\right) + \frac{1}{2}\left[{\bm \eta}^{\text{[s-d]}}\right]^T {\bm \eta}^{\text{[s-d]}},\notag
\end{align}
where ${\bm \mu} \triangleq \alpha {\bm \Lambda}_{\bm\theta^{\text{[s-d]}}} {\bm \Lambda}_{\hat{\phi}^{\text{[s-d]}}}  {\bf F} {\bf \Lambda}_{\hat{\tilde{\bf c}}} {\bf s}^{\text{[s]}}$ with ${\bm \Lambda}_{\hat{\phi}^{\text{[s-d]}}}$ and ${\bf \Lambda}_{\hat{\tilde{\bf c}}}$ are determined base on the estimated CFO and channels, $\hat{\bm\phi}^{\text{[s-d]}}$ and $\hat{\bf c}$, respectively, and ${{\bf \Sigma}}^{\text{[r]}} = \alpha^2 \sigma^2_R {\bm \Lambda}_{\bm\theta^{\text{[s-d]}}} {\bm \Lambda}_{\hat{\phi}^{\text{[s-d]}}} \hat{\bf G} \hat{\bf G}^H {\bm \Lambda}^H_{\hat{\phi}^{\text{[s-d]}}} {\bm \Lambda}^H_{\bm\theta^{\text{[s-d]}}}  + \sigma^2_D {\bf I}_N$ is the noise covariance matrix that is calculated via the estimated channels, $\hat{\bf g}$, and CFO, $\hat{\phi}^{\text{[s-d]}}$. First, the data symbols at the $k$-th iteration, $[{\bf s}^{\text{[s]}}_{\text{D}}]^{[k]}$, are used to estimate the PN at the $(k+1)$-th iteration, $[{\bm \theta}^{\text{[s-d]}}]^{[k+1]}$. To obtain a closed-form solution, as in Section \ref{sec:PN}, \eqref{V-1} is approximated by
\begin{equation}\label{V-3}\nonumber
\begin{split}
{\bf y}^{\text{[s]}} \approx & \alpha {\bm \Lambda}_{\hat{\phi}^{\text{[s-d]}}}  {\bf F} {\bf \Lambda}_{\hat{\tilde{{\bf c}}}} [{\bf s}^{\text{[s]}}]^{[k]} + {\rm Diag}\big( j\alpha {\bm \Lambda}_{\hat{\phi}^{\text{[s-d]}}}  {\bf F} {\bf \Lambda}_{\hat{\tilde{\bf c}}} [{\bf s}^{\text{[s]}}]^{[k]} \big) \\
& \times{\bf \Pi}^{\text{[s-d]}} {\bm \eta}^{\text{[s-d]}} +\alpha \hat{{\bm \Lambda}}^{[k]}_{{\bm \theta}^{\text{[s-d]}}}  {\bm \Lambda}_{\hat{\phi}^{\text{[s-d]}}}  \hat{{\bf G}} {\bf v} + {\bf w},
\end{split}
\end{equation}
where ${\bm \eta}^{\text{[s-d]}}$ denotes the shorten PN vector. By equating the gradient of \eqref{V-2} to zero, $[\hat{{\bm \theta}}^{\text{[s-d]}}]^{[k+1]}$ can be determined as\begin{equation}\label{V-4}
\begin{split}
[\hat{{\bm \theta}}^{\text{[s-d]}}]^{[k+1]} =& {\bf \Pi}^{\text{[s-d]}} \big( \Re({\bf M}^H \big[[\hat{{\bf \Sigma}}^{\text{[r]}}]^{[k]}\big]^{-1} {\bf M} ) + \frac{1}{2}{\bf I}_M \big)^{-1} \\
& \times \Re \big({\bf M}^H \big[[\hat{{\bf \Sigma}}^{\text{[r]}}]^{[k]}\big]^{-1} \\
&\;\;\;\;\;\;\;\;\;\;\;\;\;\times
({\bf y}^{\text{[s]}} - \alpha {\bm \Lambda}_{\hat{\phi}^{\text{[s-d]}}}  {\bf F} {\bf \Lambda}_{\hat{\tilde{{\bf c}}}} [{\bf s}^{\text{[s]}}]^{[k]}) \big),
\end{split}
\end{equation}
where ${\bf M} \triangleq {\rm Diag}\left( j\alpha {\bm \Lambda}_{\hat{\phi}^{\text{[s-d]}}}  {\bf F} {\bf \Lambda}_{\hat{\tilde{{\bf c}}}} [{\bf s}^{\text{[s-d]}}]^{[k]}\right) {\bf \Pi}^{\text{[s]}}$ and $[\hat{{\bf \Sigma}}^{\text{[r]}}]^{[k]} = \alpha^2 \sigma^2_R \hat{{\bm \Lambda}}^{[k]}_{\bm\theta^{\text{[s-d]}}} {\bm \Lambda}_{\hat{\phi}^{\text{[s-d]}}} \hat{\bf G} \hat{\bf G}^H {\bm \Lambda}^H_{\hat{\phi}^{\text{[s-d]}}} [\hat{{\bm \Lambda}}^{[k]}_{\bm\theta^{\text{[s-d]}}}]^H  + \sigma^2_D {\bf I}_N$. Secondly, using $[\hat{{\bm \theta}}^{\text{[s-d]}}]^{[k+1]}$ and the noise covariance matrix at the $k+1$-th iteration, $[\hat{{\bf \Sigma}}^{\text{[r]}}]^{[k+1]}$, an estimate of the transmitted symbols at the $(k+1)$-th iteration can be obtained as
\begin{align}\label{V-5}
 [{\bf s}^{\text{[s]}}_{\text{D}}]^{[k+1]} =& \big([\hat{\bf T}^{[k+1]}_{\text{D}}]^H \big[[\hat{{\bf \Sigma}}^{\text{[r]}}]^{[k+1]} \big]^{-1} \hat{\bf T}^{[k+1]}_{\text{D}} \big)^{-1} \\
& \times[\hat{\bf T}^{[k+1]}_{\text{D}}]^H \big[[\hat{{\bf \Sigma}}^{\text{[r]}}]^{[k+1]} \big]^{-1} ({\bf y}^{\text{[s]}} - \hat{\bf T}^{[k+1]}_{\text{P}} {\bf s}^{\text{[s]}}_{\text{P}} ).\notag
\end{align}
In \eqref{V-5}, although $\hat{\bf T}^{[k+1]}_{\text{P}}$ and $\hat{\bf T}^{[k+1]}_{\text{D}}$ are defined similar to ${\bf T}_{\text{P}}$ and ${\bf T}_{\text{D}}$ in \eqref{V-1}, they are obtained via the estimates  $[\hat{{\bm \theta}}^{\text{[s-d]}}]^{[k+1]}$, $\hat{\phi}^{\text{[s-d]}}$, and $\hat{\bf c}$. The overall iterative detector is given below.

\vspace{-0.2cm}
\hrulefill
\par
{\footnotesize
\textbf{Algorithm 2}
\begin{itemize}
\item \textbf{Initialize} ${\bf s}^{\text{[s]}}_{\text{D}}$ and ${{\bf \Sigma}}^{\text{[r]}}$
\item \textbf{Repeat}
\begin{itemize}
\item Update $[{\bm \theta}^{\text{[s-d]}}]^{[k+1]}$ with the estimated $[{\bf s}^{\text{[s]}}_{\text{D}}]^{[k]}$ by using \eqref{V-4} and then update $[\hat{{\bf \Sigma}}^{\text{[r]}}]^{[k]}$ as $[\hat{{\bf \Sigma}}^{\text{[r]}}]^{[k+1]}$;
\item Update $[{\bf s}^{\text{[s]}}_{\text{D}}]^{[k+1]}$ with the estimated $[{\bm \theta}^{\text{[s-d]}}]^{[k+1]}$ by using \eqref{V-5};
\end{itemize}
\item \textbf{Until} $q(n+1)-q(n) \leq \epsilon$ where $q(n)$ denotes the obtained value of objective function in \eqref{V-2} after the $n$-th iteration and $\epsilon$ is a pre-set convergence accuracy.
\end{itemize}}
\vspace{-0.3cm}
\hrulefill
\begin{figure*}[t]
\centering
  \begin{minipage}[t]{0.48\textwidth}
  \centering
     \hspace{-.5cm}\includegraphics[scale=0.54]{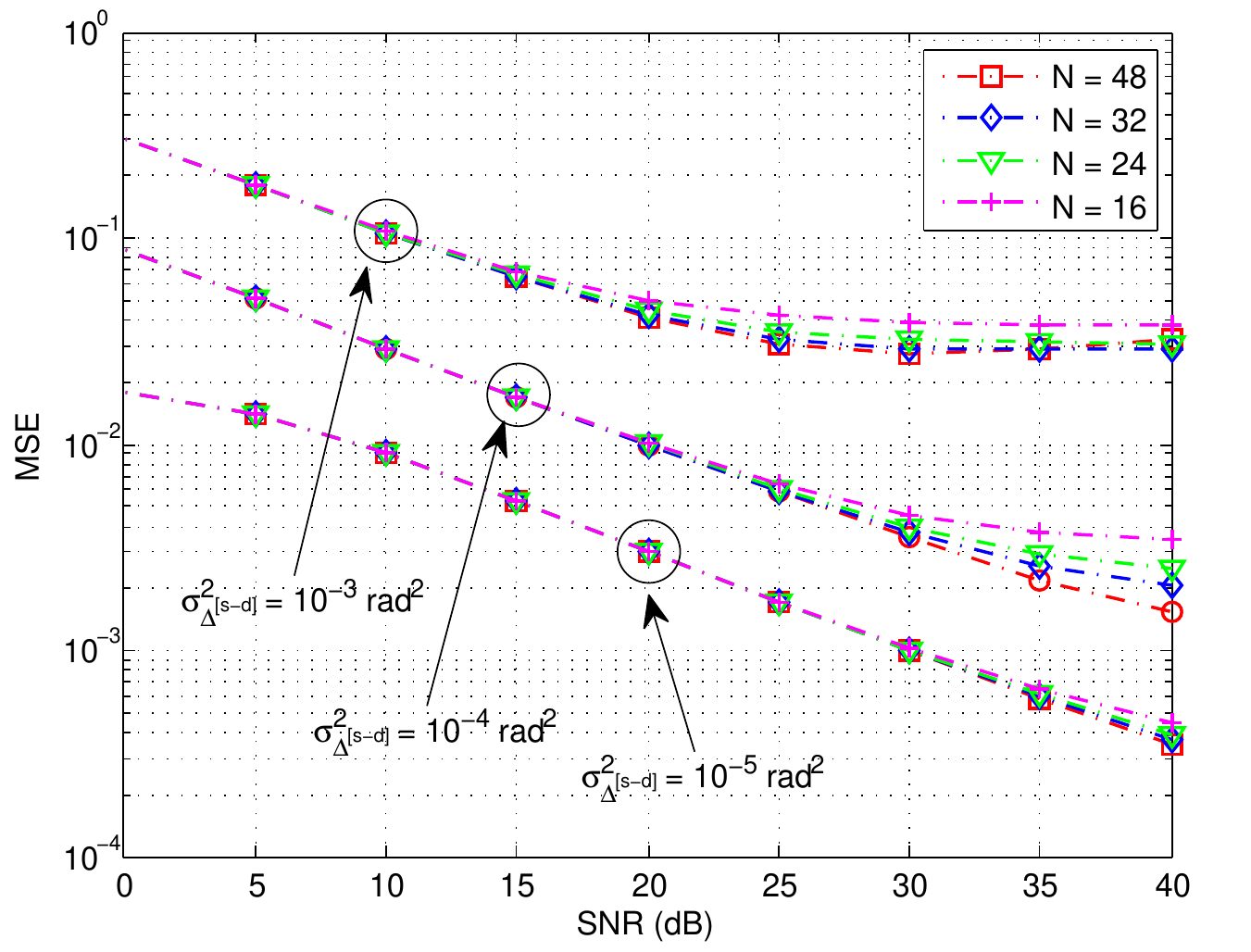}
\vspace{-0.4cm}
\caption{The MSE of phase noise estimation with different $M$.} \label{Fig_PN}
  \end{minipage}
  \begin{minipage}[t]{0.48\textwidth}
  \centering
    \hspace{-.2cm}\includegraphics[scale=0.54]{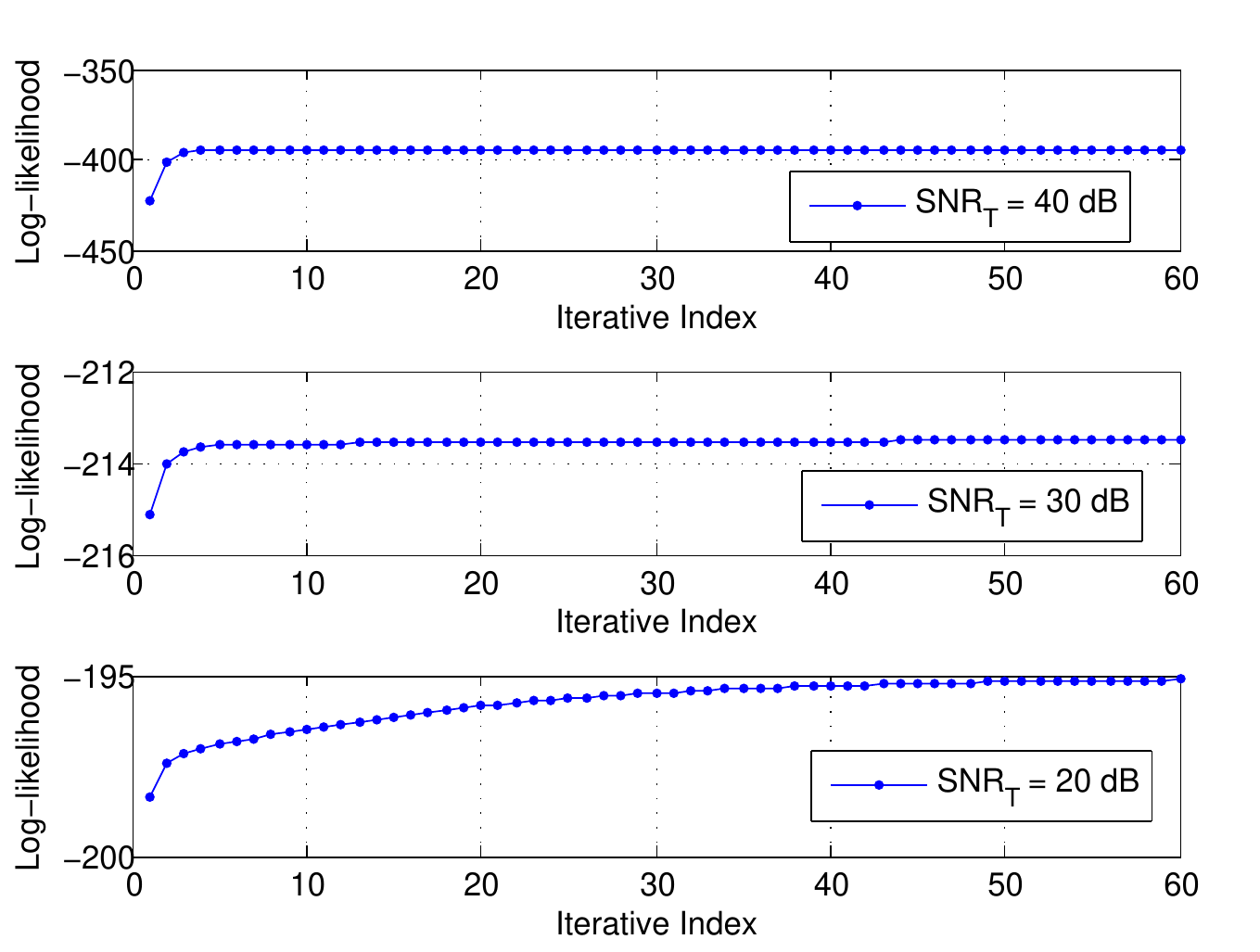}
    \vspace{-0.4cm}
    \caption{The convergence of the proposed joint estimation algorithm at different SNR ($\sigma^2_{\Delta^{\text{[s-d]}}}=10^{-4}~ {\rm rad}^2$).} \label{Iter_Num}
  \end{minipage}
  \vspace*{-.5cm}
\end{figure*}

In Algorithm 2, initial estimates of ${\bf s}^{\text{[s]}}_{\text{D}}$ and ${{\bf \Sigma}}^{\text{[r]}}$ are obtained similar to that of the training interval.

\begin{remark}\label{rem:1} As indicate here, the ambiguities associated with calculating the MSE for channel response, CFO, and PN parameters do not affect the data transmission interval. Let us denote the ambiguities of the channels and CFO in the training phase as
$\hat{{\bf g}} \rightarrow \exp(-j\varphi_g){\bf g}$,
$\hat{{\bf c}} \rightarrow \exp(-j(\varphi_h+\varphi_g)){\bf c}$ and
$\hat{\phi}^{\text{[s-d]}} \rightarrow \phi^{\text{[s-d]}} - \epsilon^{\text{[s-d]}}$. These ambiguities can be combined during the data transmission phase in the overall estimate of the PN parameters ${\bm \theta}^{\text{[s-d]}}$ in \eqref{V-1}, which can be written as $\hat{{\bm \theta}}^{\text{[s-d]}} \rightarrow {\bm \theta}^{\text{[s-d]}} + (\varphi_g+\varphi_h) {\bf 1} + {\bm \epsilon}^{\text{[s-d]}}$ (${\bm \epsilon}^{\text{[s-d]}}$ is defined in \eqref{III-28}). It can be clearly observed that these ambiguities do not affect the overall channel response, $\alpha {\bm \Lambda}_{\bm\theta^{\text{[s-d]}}} {\bm \Lambda}_{\phi^{\text{[s-d]}}}  {\bf F} {\bf \Lambda}_{\tilde{\bf c}}$, and the received signal in \eqref{V-1}.

\end{remark}

\section{Simulation Results}\label{sec:simulations}

In this section, extensive simulations are carried out to evaluate the performance of the proposed algorithms.
In all the simulations, it is assumed that the multi-path channels exhibit unit-variance Rayleigh fading characteristics. Without loss of generality, it is assumed that the noise powers at relay and destination nodes are the same, i.e., $\sigma^2_R = \sigma^2_D =1$. Moreover, the following simulation parameters are considered:
\begin{itemize}
  \item [1)] The multipath fading channels from relay-to-destination and source-to-relay, ${\bf g}$ and ${\bf h}$, respectively, are assumed to consist of $6$ taps, i.e., $L_g=L_h=6$,
  \item [2)] $N=64$ subcarriers are used in each OFDM symbol and all the subcarriers are modulated in quadrature phase shift keying (QPSK) format for both training and data transmission phases,
  \item [3)] The normalized CFOs, $\phi^{\text{[s-d]}}$ and $\phi^{\text{[r-d]}}$, are uniformly drawn from $[-0.4, 0.4]$ and $[-0.2, 0.2]$, respectively, and
  \item [4)] The PN innovation variances for, ${\bm \theta}^{\text{[s-d]}}$ and  ${\bm \theta}^{\text{[r-d]}}$ are assumed to be the same, i.e., ${\sigma^2_{\Delta^{\text{[s-d]}}}} = {\sigma^2_{\Delta^{\text{[r-d]}}}} = \sigma^2_{\Delta}$.
\end{itemize}
Let us outline the choice of the scaling factor at the relay here. After removing the CP, the received signal vector at the relay in the frequency domain, ${\bf z}$, is given by
\begin{align*}
{\bf z} \triangleq \tilde{{\bf h}} \odot {\bf s} + {\bf n} \in \mathbb{C}^{N \times 1},
\end{align*}
where $\tilde{{\bf h}} \triangleq  [\tilde{{h}}_1, \tilde{{h}}_2,\cdots,\tilde{{h}}_{N}]^T$ with
$\tilde{{h}}_k = \sum^{L_h-1}_{n=0} \exp{\frac{-j2\pi kn}{K}} h(n)$, for $k=0,\cdots,N-1$. In addition, it is assumed that $\tilde{{h}}_k\sim{\cal CN}(0, L_h \sigma^2_h)$ with $\sigma^2_h$ denoting the variance of $h(n)$. Hence, the received signal power, $P_{\bf z}$, is given by $P_{\bf z} = \mathbb{E}_{{\bf h},{\bf s}}(||{\bf z}||^2_2)=NL_h \sigma^2_h P^{\text{[s]}}_{\text{[T]}} + N \sigma^2_R$. By considering the added CP at the source, the total power of the received signal at the relay node can be approximated as
$\bar{P}_{\bf z} = P_{\bf z}\frac{N_{CP}+N}{N}$. Thus, the relay scaling factor, $\alpha$, can be determined as ${\displaystyle\alpha = \sqrt{ {P^{\text{[r]}}_{\text{T}}}/ {\bar{P}_{\bf z}}}}$. Without loss of generality, in the remainder of this section, $\alpha =1$ by letting $P^{\text{[r]}}_{\text{T}}=\bar{P}_{\bf z}$. Moreover, it is assumed that $P^{\text{[s]}}_{\text{T}} = P^{\text{[r]}}_{\text{T}}=P_{\text{T}}=$SNR.\footnote{Due to lack of any prior art on the impact of PN on relaying networks, the performance of the proposed estimator and receiver structure cannot be compared with any existing algorithms.}

\begin{figure*}[t]
\centering
  \begin{minipage}[t]{0.48\textwidth}
  \centering
     \hspace{-.5cm}\includegraphics[scale=0.54]{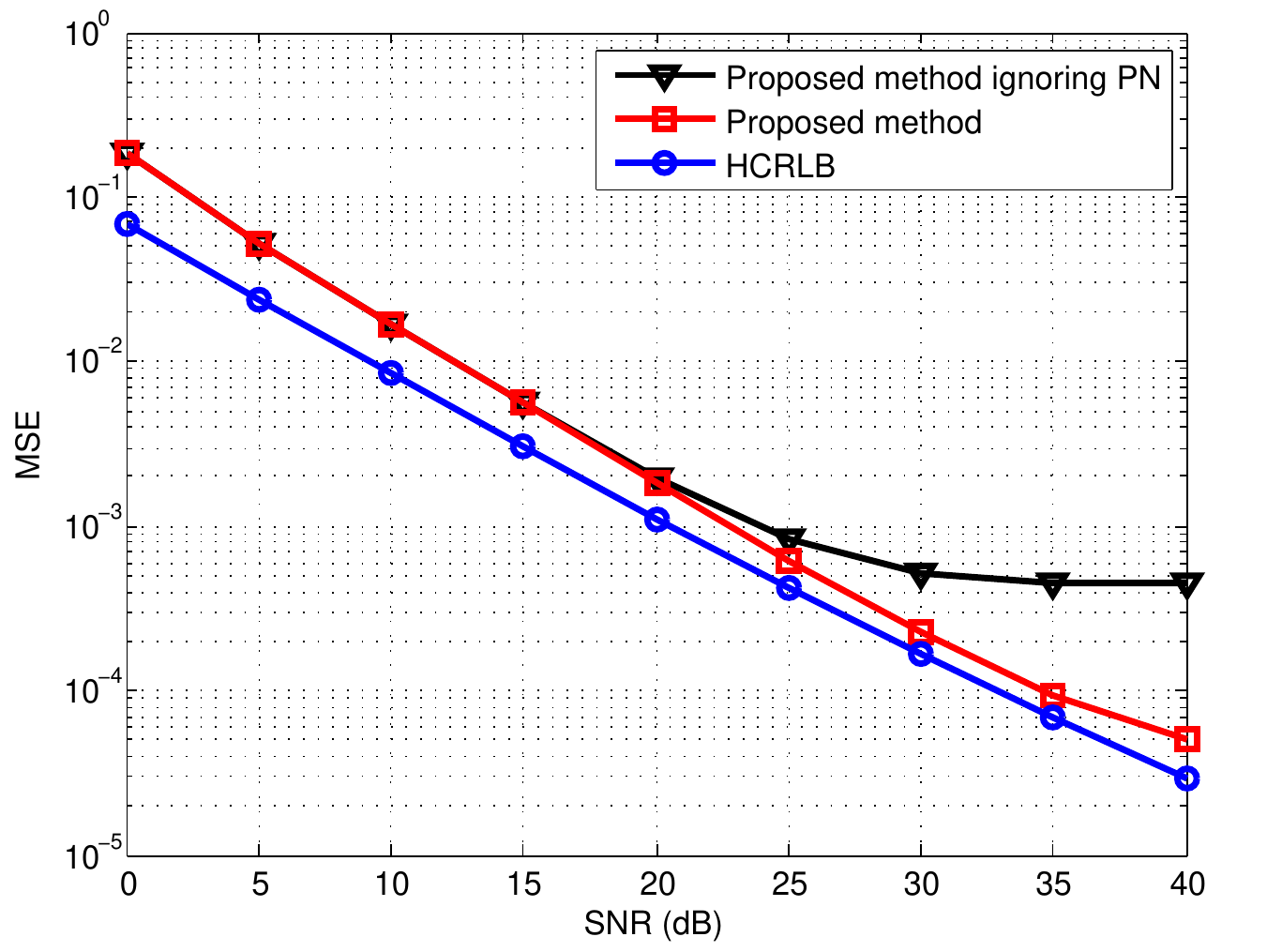}
\vspace{-0.5cm}
\caption{The MSE of $\underline{{\bf g}}$ estimation at $\sigma^2_{\Delta} = 10^{-4} {\rm rad}^2$.} \label{Channel_g_e_4}
  \end{minipage}
  \begin{minipage}[t]{0.48\textwidth}
  \centering
    \hspace{-.2cm}\includegraphics[scale=0.54]{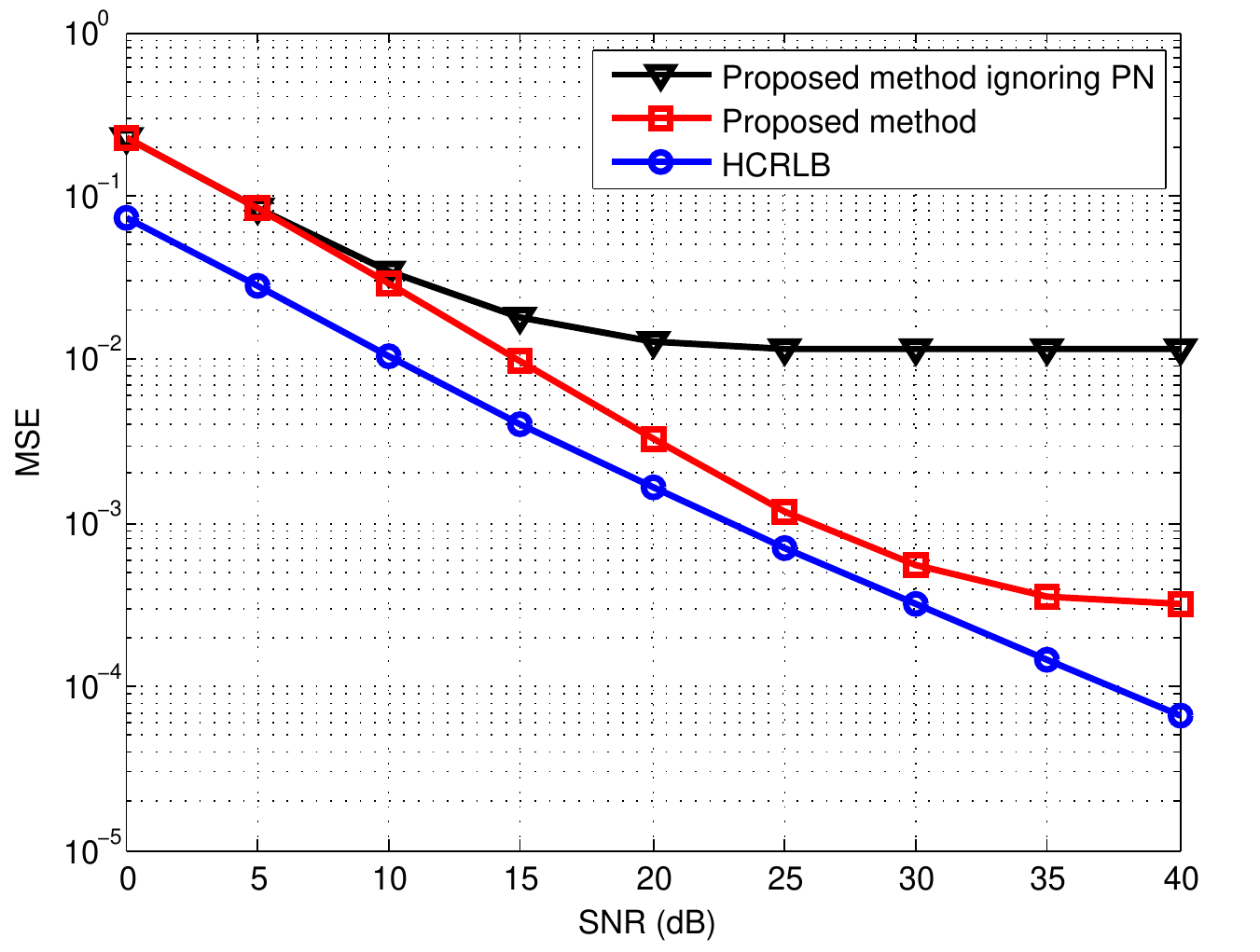}
    \vspace{-0.5cm}
    \caption{The MSE of $\underline{{\bf g}}$ estimation at $\sigma^2_{\Delta} = 10^{-3} {\rm rad}^2$.} \label{Channel_g_e_3}
  \end{minipage}
  \vspace*{-.3cm}
\end{figure*}

Fig.~\ref{Fig_PN} depicts the MSE of PN estimation, when estimating the shortened phase vector ${\bm \eta}^{\text{[s-d]}}$ for different values of $M$ (see Section \ref{sec:short-PN}). For ease of comparison and to isolate the effect of CFO and channel estimation, it is assumed that the channel response, ${\bf c}$, and the CFO, $\phi^{\text{[s-d]}}$, are perfectly known. From the plots in Fig.~\ref{Fig_PN} it can be concluded that when the PN innovation variance $\sigma^2_{\Delta^{\text{[s-d]}}}$ is small, i.e., $\sigma^2_{\Delta^{\text{[s-d]}}}=10^{-5}~ {\rm rad}^2$, PN parameters can be accurately estimated using $M=16$ compared to $M=64$. Such an approach greatly reduces the PN estimation overhead. For scenarios with higher innovation variances, i.e., $\sigma^2_{\Delta^{\text{[s-d]}}}=10^{-4}~ {\rm rad}^2$ and $\sigma^2_{\Delta^{\text{[s-d]}}}=10^{-3}~ {\rm rad}^2$, it can be deduced that a larger value of $M$ is needed to ensure accurate PN estimation, e.g., $M=32$. However, even for these larger PN variances, using the proposed scheme, the number of PN parameters that need to be tracked is reduced by one half. Accordingly, in the remainder of this section, $M=32$.


%


In Fig.~\ref{Iter_Num}, the convergence of the proposed joint estimation algorithm is plotted for different SNRs. It can be observed that on average less than $50$ iterations are needed for the proposed algorithm to coverage to the true estimates for a wide range of SNR values. More importantly, the result in Fig.~\ref{Iter_Num} show that as the SNR increases the proposed algorithm converges more quickly, e.g., for SNR$=30$ dB less than $10$ iterations are needed for the proposed estimator to converge.

\begin{figure*}[b]
\centering
  \begin{minipage}[t]{0.48\textwidth}
  \centering
     \hspace{-.5cm}\includegraphics[scale=0.54]{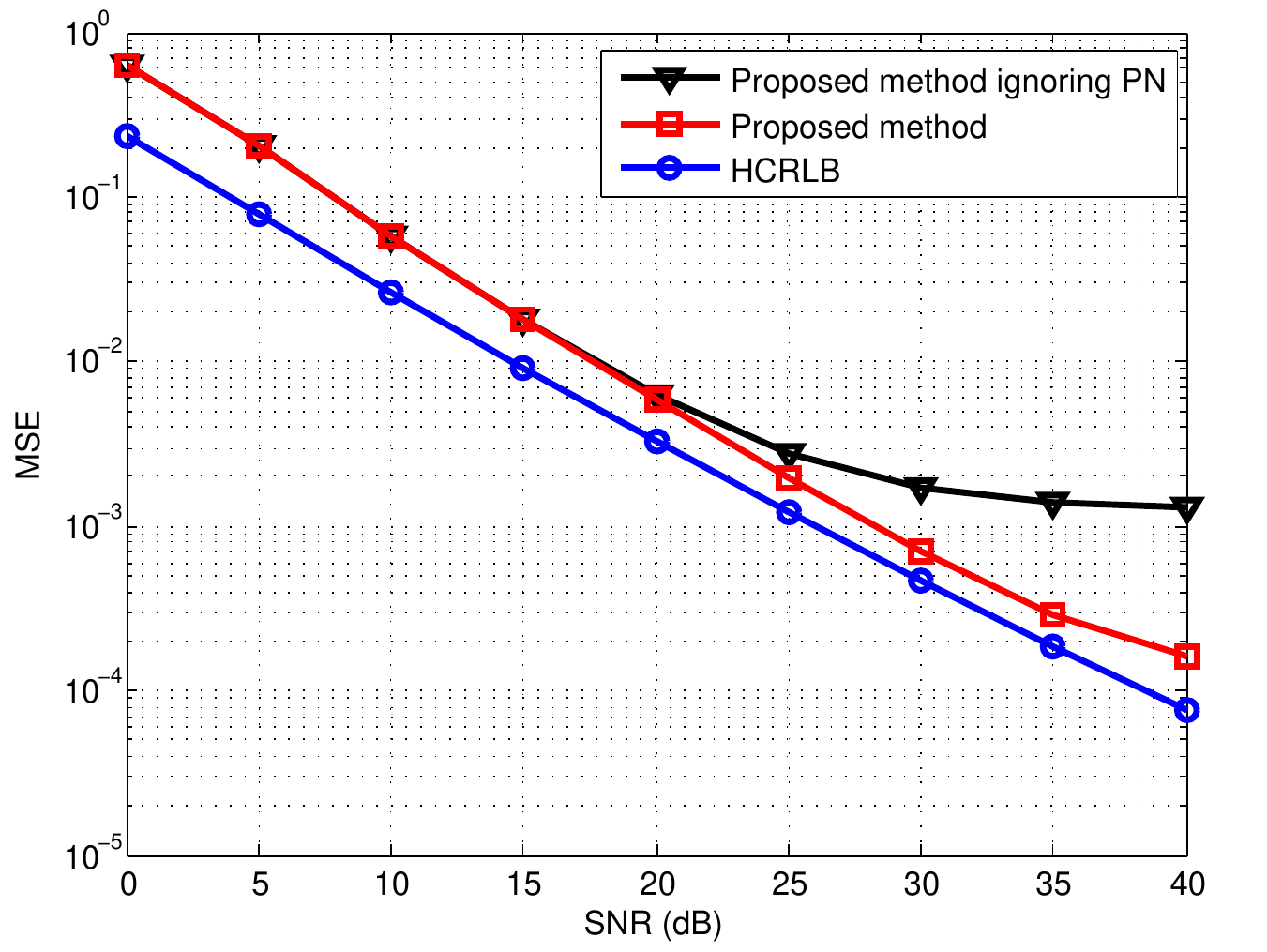}
\vspace{-0.5cm}
    \caption{The MSE of $\underline{{\bf h}}$ estimation at $\sigma^2_{\Delta} = 10^{-4} {\rm rad}^2$.} \label{Channel_h_e_4}
  \end{minipage}
  \begin{minipage}[t]{0.48\textwidth}
  \centering
    \hspace{-.2cm}\includegraphics[scale=0.54]{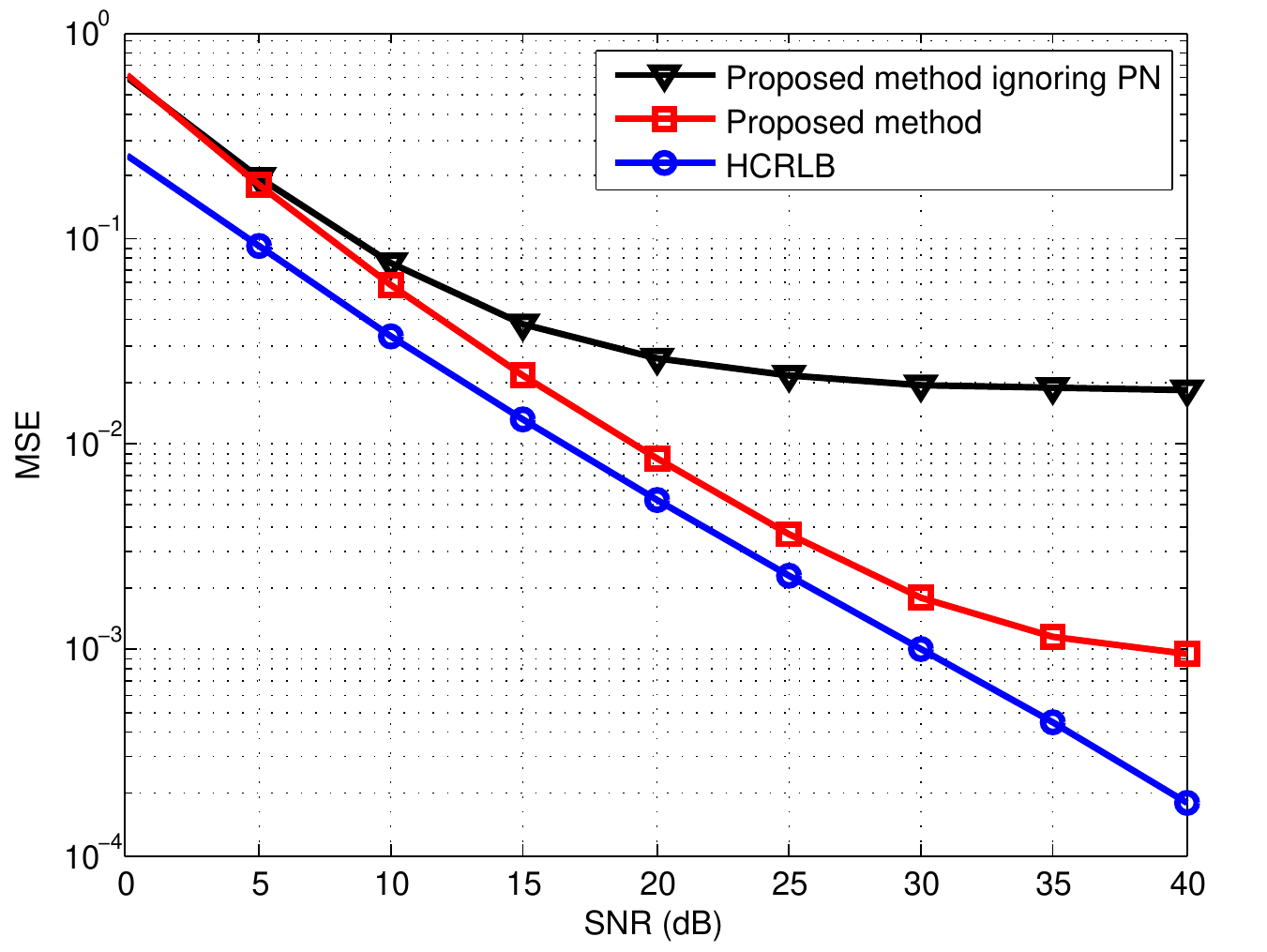}
    \vspace{-0.5cm}
    \caption{The MSE of $\underline{{\bf h}}$ estimation at $\sigma^2_{\Delta} = 10^{-3} {\rm rad}^2$.} \label{Channel_h_e_3}
  \end{minipage}
  \vspace*{-.5cm}
\end{figure*}

Figs.~\ref{Channel_g_e_4} and \ref{Channel_g_e_3} illustrate the estimation MSE of relay-to-destination channel, $\underline{{\bf g}}$ (defined in \emph{Remark \ref{rem:0}}), for PN variances, $\sigma^2_{\Delta}=10^{-4}~ {\rm rad}^2$ and $\sigma^2_{\Delta}=10^{-3}~ {\rm rad}^2$, respectively, while the estimation MSE of the source-to-relay channel, $\underline{{\bf h}}$ (defined in \emph{Remark \ref{rem:0}}), is presented in Figs.~\ref{Channel_h_e_4} and ~\ref{Channel_h_e_3}. As a comparison, the channel estimation performance while ignoring the effect of PN on the received signal is also presented in these figures. Finally, the proposed estimation algorithms performance is benchmarked using the derived HCRLB in Section \ref{sec:HCRLB}. Figs.~\ref{Channel_g_e_4}--\ref{Channel_h_e_3} indicate that by including the PN parameters in the joint estimation problem, channel estimation performance in relay networks can be significantly enhanced. At moderate SNR, Figs.~\ref{Channel_g_e_4}--\ref{Channel_h_e_3} also show that the proposed algorithm has a constant performance gap with respect to the derived HCRLB bound for both PN innovation variances of $\sigma^2_{\Delta} = 10^{-4}~ {\rm rad}^2$ and $\sigma^2_{\Delta} = 10^{-3}~ {\rm rad}^2$. This is due to the inherent structure of the HCRLB, which is not necessarily a very tight bound as stated in \cite{Trees2007}. Nevertheless, the performance of the proposed estimator is close to the derived HCRLB for moderate SNR. Finally, the results in Figs.~\ref{Channel_g_e_4}--\ref{Channel_h_e_3} indicate that for large PN innovation variances, e.g., $\sigma^2_{\Delta^{\text{[s-d]}}} = 10^{-3}~ {\rm rad}^2$, the channel estimation performance suffers from an MSE error-floor at high SNR. This error-floor is caused by the time-varying PN parameters that cannot be perfectly estimated. Hence, at low SNR the overall estimation performance of the estimator is limited by the additive noise at the destination node, while at high SNR the algorithm's estimation performance is limited by the PN.

\begin{figure*}[t]
\centering
  \begin{minipage}[t]{0.48\textwidth}
  \centering
     \hspace{-.5cm}\includegraphics[scale=0.54]{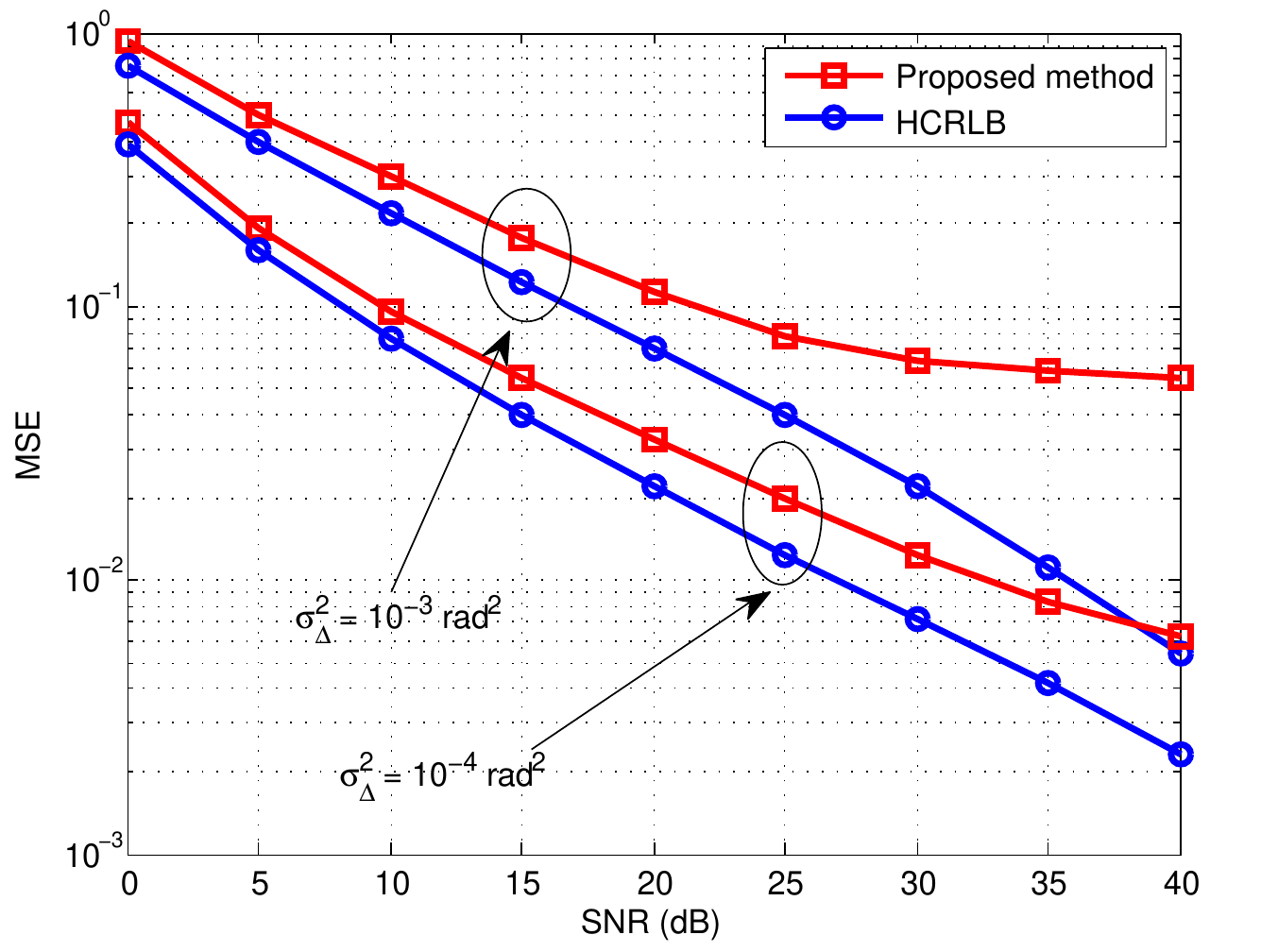}
    \vspace{-0.5cm}
    \caption{The MSE of CFO plus PN estimation at $\sigma^2_{\Delta}=10^{-3} {\rm rad}^2 $ \newline and $\sigma^2_{\Delta}=10^{-4} {\rm rad}^2 $.} \label{CFO_PN_e_3}
    \label{CFO_PN_e_4}
  \end{minipage}
  \begin{minipage}[t]{0.48\textwidth}
  \centering
    \hspace{-.2cm}\includegraphics[scale=0.54]{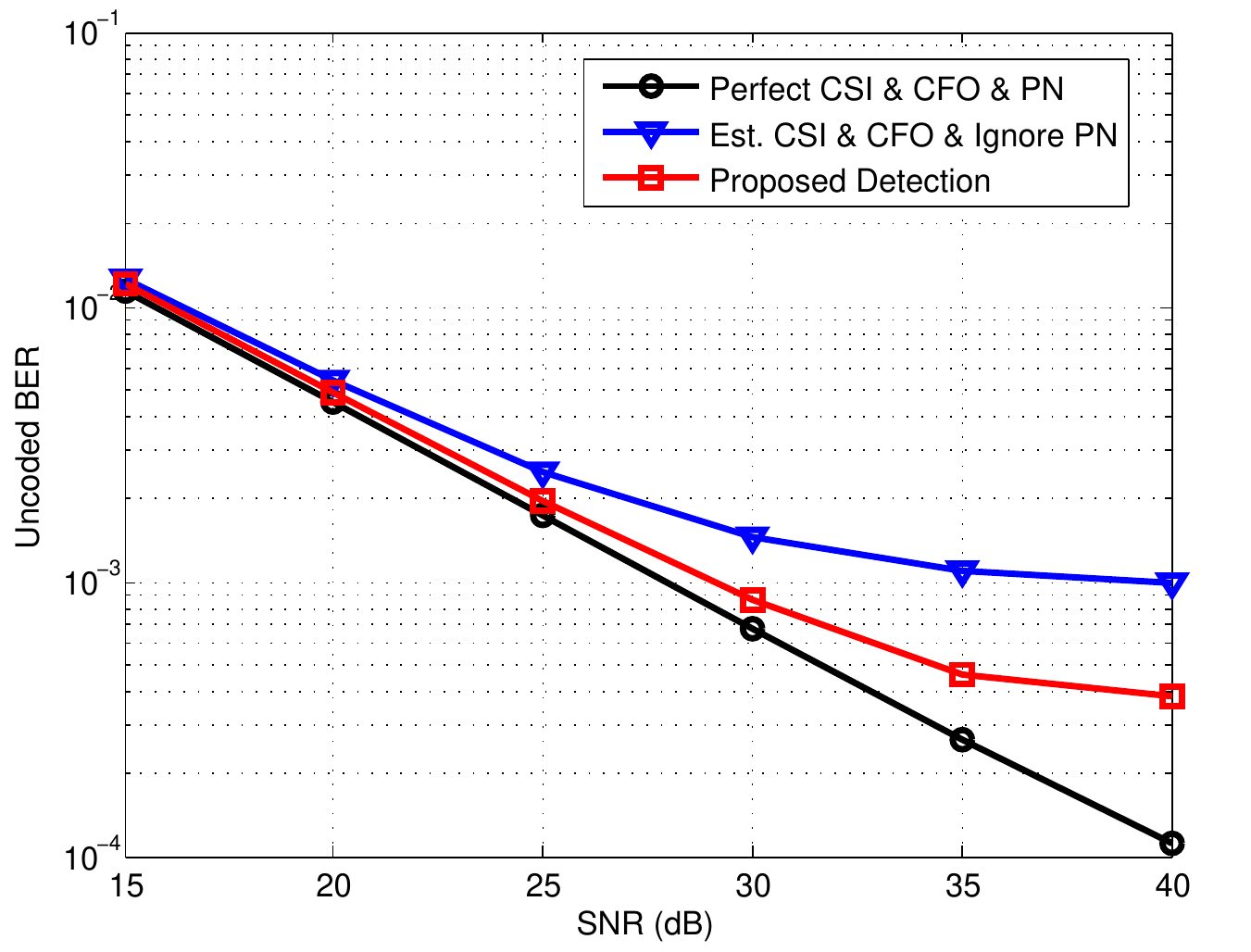}
\vspace{-0.5cm}
    \caption{The BER performance for the proposed joint data detection algorithm at $\sigma^2_{\Delta}=10^{-4} {\rm rad}^2$.} \label{BER_e_4}
  \end{minipage}
  \vspace*{-0.5cm}
\end{figure*}

Fig. \ref{CFO_PN_e_3} illustrates the MSE for estimation of combined CFO and PN, $\underline{{\bm \delta}}$ for different PN variances. Similar to the results for channel estimation, the overall estimation performance suffers from an error floor for large PN variances, e.g., $\sigma^2_{\Delta} = 10^{-3}~ {\rm rad}^2$. This phenomenon can be similarly justified due to the imperfect estimation of PN parameters. Moreover, there is a $5$ dB gap between the CFO and PN estimation MSE and the derived HCRLB at medium SNRs.


Fig.~\ref{BER_e_4} illustrates the end-to-end BER of an uncoded OFDM relay network when applying the combination of the proposed iterative estimator and detector at $\sigma^2_{\Delta}=10^{-4} {\rm rad}^2$. It is observed that significant performance gains can be achieved by using the proposed joint data detection and PN estimation algorithm compared to a scheme that ignores the impact of PN. However, compared to the case with perfect channel, CFO, and PN, the proposed data detection algorithm still suffers from an error-floor at high SNR regime. This can be again attributed to imperfect PN estimation, where at high SNR, the overall BER of the OFDM relay system is dominated by PN and not the additive noise. This result indicates the importance of considering the impact of PN when determining the link budget, throughput, and coverage of wireless relay networks.

\vspace{-0pt}
\section{Conclusions}
In this paper, joint channel, CFO, and PN estimation and data detection in OFDM relay networks is analyzed. Due to its time-varying nature, new algorithms for tracking the PN parameters in both the training and data transmission intervals are proposed. During the training interval, a new joint CFO, channel, and PN estimation algorithm that iteratively estimates these impairments is derived. To reduce estimation overhead, the proposed algorithm applies the correlation amongst the PN parameters to reduce the dimensionality of the estimation problem. Simulations show that the proposed estimator significantly enhances channel estimation performance in presence of PN, converges quickly, and performs close to the derived HCRLB at medium SNRs. Moreover, an iterative joint PN estimation and data detection receiver based on the MAP criterion at the destination node is proposed. The combination of the proposed estimation and data detection algorithms is shown to result in $5$--$10$ dB performance gains over schemes that ignore the deteriorating effect of PN. 



\vspace{-0pt}

\appendices
\numberwithin{equation}{section}
\vspace{-12pt}
\section{Derivation of \eqref{III-10}}
\label{prof_PN}
In this section an expression for the optimization in \eqref{III-7} is derived. It is straightforward to determine that the optimization in \eqref{III-7} is a nonlinear and non-convex problem. Thus, the solution of ${\bm \eta}^{\text{[s-d]}}$ in \eqref{III-7} should be in general obtained through exhaustive search. To simplify the problem and obtain a closed-form solution, we first approximate the covariance matrix ${\bf \Sigma}^{\text{[r]}}$ as
$[\hat{{\bf \Sigma}}^{\text{[r]}}]^{[k]}=\alpha^2 \sigma^2_R \hat{{\bm \Lambda}}^{[k]}_{\bm\theta^{\text{[s-d]}}} \hat{{\bm \Lambda}}^{[k]}_{\phi^{\text{[s-d]}}} \hat{\bf G}^{[k]} [\hat{\bf G}^{[k]}]^H  [\hat{{\bm \Lambda}}^{[k]}_{\phi^{\text{[s-d]}}}]^H [\hat{{\bm \Lambda}}^{[k]}_{\bm\theta^{\text{[s-d]}}}]^H + \sigma^2_D {\bf I}_N$,
where $[\hat{{\bm \Lambda}}^{[k]}_{\bm\theta^{\text{[s-d]}}}]_{m,m}= e^{j[\hat{\theta}^{\text{[s-d]}}(m)]^{[k]}}$ is obtained from the previous iteration. Moreover, since the PN innovation variance of practical oscillators is usually small, the elements in ${\bm \Lambda}_{\bm\theta^{\text{[s-d]}}}$ can be approximated by a Taylor series expansion as $e^{j\theta^{\text{[s-d]}}(n)} \approx  1+j\theta^{\text{[s-d]}}(n)$. This small angle approximation has also been used in \cite{Darryl2006, Jun2009, Mehrpouyan2012} for PN estimation. Hence, the PN matrix, ${\bm \Lambda}_{\bm\theta^{\text{[s-d]}}}$, can be approximated as ${\bm \Lambda}_{\bm\theta^{\text{[s-d]}}} \approx {\bf I}_N +j {\rm Diag}({\bm \theta}^{\text{[s-d]}})$ and $\mathcal{L}_{{\bm \eta}^{\text{[s-d]}}}$ in \eqref{III-7} can be rewritten as
\begin{equation}\label{III-8}
\begin{split}
  \mathcal{L}_{{\bm \eta}^{\text{[s-d]}}} \approx &
   (\bar{{\bf y}}^{\text{[s]}} - {\bf B} {\bm \eta}^{\text{[s-d]}})^H \left[[\hat{{\bf \Sigma}}^{\text{[r]}}]^{[k]} \right]^{-1} (\bar{{\bf y}}^{\text{[s]}} - {\bf B} {\bm \eta}^{\text{[s-d]}}) \\
 &  + \frac{1}{2}[{\bm \eta}^{\text{[s-d]}}]^T {\bm \eta}^{\text{[s-d]}} \\
\approx &~ [\bar{{\bf y}}^{\text{[s]}}]^H \left[[\hat{{\bf \Sigma}}^{\text{[r]}}]^{[k]} \right]^{-1} \bar{{\bf y}}^{\text{[s]}} - 2 \Re([\bar{{\bf y}}^{\text{[s]}}]^H \left[[\hat{{\bf \Sigma}}^{\text{[r]}}]^{[k]} \right]^{-1} {\bf B}) \\
 & \times {\bm \eta}^{\text{[s-d]}} + [{\bm \eta}^{\text{[s-d]}}]^T \Re({\bf B}^H \left[[\hat{{\bf \Sigma}}^{\text{[r]}}]^{[k]} \right]^{-1} {\bf B}){\bm \eta}^{\text{[s-d]}} \\
 & + \frac{1}{2}[{\bm \eta}^{\text{[s-d]}}]^T {\bm \eta}^{\text{[s-d]}},
\end{split}
\end{equation}
where $\bar{{\bf y}}^{\text{[s]}} \triangleq {\bf y}^{\text{[s]}} -  \alpha \hat{\bm \Lambda}^{[k]}_{\phi^{\text{[s-d]}}}  {\bf F}^H {\bm \Lambda}_{s^{\text{[s]}}} {\bf F}_{[L]}\hat{{\bf c}}^{[k]}$ and ${\bf B} \triangleq j{\rm Diag}( \alpha \hat{\bm \Lambda}^{[k]}_{\phi^{\text{[s-d]}}}  {\bf F}^H {\bm \Lambda}_{s^{\text{[s]}}} {\bf F}_{[L]}\hat{{\bf c}}^{[k]}) {\bf \Pi}^{\text{[s-d]}}$. Next, by equating the gradient of \eqref{III-8} to zero, i.e.,
\begin{equation}\label{III-9} \nonumber
\begin{split}
 \frac{\partial \mathcal{L}_{{\bm \eta}^{\text{[s-d]}}}}{\partial {\bm \eta}^{\text{[s-d]}}} =& -2 \Re({\bf B}^H \left[[\hat{{\bf \Sigma}}^{\text{[r]}}]^{[k]} \right]^{-1} \bar{{\bf y}}^{\text{[s]}} )   \\
& +2  \Re({\bf B}^H \left[[\hat{{\bf \Sigma}}^{\text{[r]}}]^{[k]} \right]^{-1} {\bf B}){\bm \eta}^{\text{[s-d]}} +{\bm \eta}^{\text{[s-d]}} \\
=& {\bf 0}_{M\times 1},
\end{split}
\end{equation}
Then we obtain \eqref{III-10}.

\numberwithin{equation}{section}
\section{Derivation of FIM}
\label{prof_Theorem1}

In this section, the FIM for joint estimation of channels, CFO, and PN parameters, i.e., ${\bf FIM}({\bf y};{\bm \lambda})$, is derived. First, note that the combined received signal vector at the destination node in \eqref{III-33}, ${\bf y} \triangleq \left[ [{\bf y}^{\text{[s]}}]^T, [{\bf y}^{\text{[r]}}]^T \right]^T$ is a multivariate Gaussian random variable, i.e., ${\bf y}\thicksim {\cal N}(\underline{{\bm \mu}}, \underline{{\bf \Sigma}}) $ with mean $\underline{{\bm \mu}} = \big[(\alpha {\bm \Lambda}_{\bm\theta^{\text{[s-d]}}} {\bm \Lambda}_{\phi^{\text{[s-d]}}}  {\bf F}^H \underline{{\bm \Lambda}}_{s^{\text{[s]}}} {\bf F}_{[L]}\underline{{\bf c}})^T, ({\bm \Lambda}_{\theta^{\text{[r-d]}}} {\bm \Lambda}_{\phi^{\text{[r-d]}}}  {\bf F}^H$ $ \underline{{\bm \Lambda}}_{s^{\text{[r]}}} {\bf F}_{[L_g]}\underline{{\bf g}})^T \big]^T$ and covariance $\underline{{\bf \Sigma}} = {\rm Blkdiag}(\underline{{\bf \Sigma}}^{\text{[r]}}, \sigma^2_D {\bf I}_N)$. As a result, the $(i,j)$-th element of ${\bf FIM}({\bf y};{\bm \lambda})$ can be determined as \cite{StevenBook}\vspace{-4pt}
\begin{equation}\label{III-36}
\begin{split}
[{\bf FIM}({\bf y};{\bm \lambda})]_{i,j} = & 2 {\rm Re} \Big[\frac{\partial \underline{{\bm \mu}}^H}{\partial \lambda_i} \underline{{\bf \Sigma}}^{-1} \frac{\partial \underline{{\bm \mu}}}{\partial \lambda_j} \Big]  \\
&+ {\rm Tr}\Big[ \underline{{\bf \Sigma}}^{-1} \frac{\partial \underline{{\bf \Sigma}}}{\partial \lambda_i} \underline{{\bf \Sigma}}^{-1} \frac{\partial \underline{{\bf \Sigma}}}{\partial \lambda_j} \Big].
\end{split}
\end{equation}
To obtain \eqref{III-36}, the following derivatives are evaluated as
\begin{equation}\label{III-36-1}
\begin{split}
\frac{\partial \underline{{\bm \mu}}}{\partial \phi^{\text{[s-d]}}} &= \left[
                                                             (\alpha{\bm \Lambda} {\bm \Lambda}_{\bm\theta^{\text{[s-d]}}} {\bm \Lambda}_{\bm\phi^{\text{[s-d]}}}  {\bf F}^H \underline{{\bm \Lambda}}_{s^{\text{[s]}}} {\bf F}_{[L]}\underline{{\bf c}})^T,
                                                             {\bf 0}^T_{N \times 1}
                                                         \right]^T,\\
\frac{\partial \underline{{\bm \mu}}}{\partial \phi^{\text{[r-d]}}} &= \left[
                                                             {\bf 0}^T_{N \times 1},
                                                             ({\bm \Lambda} {\bm \Lambda}_{\theta^{\text{[r-d]}}} {\bm \Lambda}_{\phi^{\text{[r-d]}}}  {\bf F}^H \underline{{\bm \Lambda}}_{s^{\text{[r]}}} {\bf F}_{[L_g]}\underline{{\bf g}})^T                                                          
                                                         \right]^T,\\
\frac{\partial \underline{{\bm \mu}}}{\partial \theta^{\text{[s-d]}} (m)} &= \left[
                                                            ( {\rm Diag} \left(\alpha {\bm \Lambda}_{\phi^{\text{[s-d]}}}  {\bf F}^H \underline{{\bm \Lambda}}_{s^{\text{[s]}}} {\bf F}_{[L]}\underline{{\bf c}}\right){\bf a}_{m} )^T,
                                                             {\bf 0}^T_{N \times 1}
                                                         \right]^T,\\
                                                         \frac{\partial \underline{{\bm \mu}}}{\partial \theta^{\text{[r-d]}}(m)} &= \left[
                                                             {\bf 0}^T_{N \times 1},
                                                             ( {\rm Diag} \left( {\bm \Lambda}_{\phi^{\text{[r-d]}}}  {\bf F}^H \underline{{\bm \Lambda}}_{s^{\text{[r]}}} {\bf F}_{[L_g]}\underline{{\bf g}}\right){\bf b}_{m})^T
                                                         \right]^T,
\end{split}
\end{equation}
where ${\bf a}_{m} $ and ${\bf b}_{m} $ are defined below \eqref{III-Add-2}. Moreover, for channel responses $\underline{{\bf g}}$ and $\underline{{\bf h}}$, for $m=1,2,\cdots,L_g$, we have $\frac{\partial \underline{{\bm \mu}}}{\partial ({g(0)})} = {\bf E}(:,1)$ and
\begin{equation}\label{III-37}
\begin{split}
\frac{\partial \underline{{\bm \mu}}}{\partial \Re({g(m)})} = {\bf E}(:,m),~
 \frac{\partial \underline{{\bm \mu}}}{\partial \Im({g(m)})} &= j{\bf E}(:,m),
\end{split}
\end{equation}
and, for $m=1,2,\cdots,L_h$, we have $\frac{\partial \underline{{\bm \mu}}}{\partial ({h(0)})} = {\bf K}(:,1)$ and
\begin{equation}\label{III-37-1}
\begin{split}
\frac{\partial \underline{{\bm \mu}}}{\partial \Re({h(m)})} = {\bf K}(:,m), ~
\frac{\partial \underline{{\bm \mu}}}{\partial \Im({h(m)})} &= j{\bf K}(:,m),
\end{split}
\end{equation}
where ${\bf E} $ and ${\bf K} $ are defined as in \eqref{III-Add-2-1} and \eqref{III-Add-2_5}, respectively. Since $\phi^{\text{[r-d]}}$, ${\bm \theta}^{\text{[r-d]}}$, $\underline{{\bf h}}$ are irrelevant to the noise covariance matrix ${\bf \Sigma}$, it is straightforward to determine that
$\frac{\partial \underline{{\bf \Sigma}}}{\partial \phi^{\text{[r-d]}}}  = \frac{\partial \underline{{\bf \Sigma}}}{\partial \theta^{\text{[r-d]}}(m)}
=\frac{\partial
\underline{{\bf \Sigma}}}{\partial {\underline{h}(0)}}
=\frac{\partial
\underline{{\bf \Sigma}}}{\partial \Re({\underline{h}(m)})} = \frac{\partial \underline{{\bf \Sigma}}}{\partial \Im({\underline{h}(m)})} = {\bf 0}$, $\forall m$.
Moreover, for the CFO and PN parameters, $\phi^{\text{[s-d]}}$ and ${\bm \theta}^{\text{[s-d]}}$, we can obtain that $\frac{\partial \underline{{\bf \Sigma}}}{\partial \phi^{\text{[s-d]}}}={\rm Blkdiag} \big( \frac{\partial \underline{{\bf \Sigma}}^{\text{[r]}}}{\partial \phi^{\text{[s-d]}}}, {\bf 0}_{N\times N} \big) $, where
$\frac{\partial \underline{{\bf \Sigma}}^{\text{[r]}}}{\partial \phi^{\text{[s-d]}}} = \big(\alpha^2 \sigma^2_R {\bm \Lambda}_{\bm\theta^{\text{[s-d]}}} \underline{{\bf G}}\underline{{\bf G}}^H  {\bm \Lambda}^H_{\bm\theta^{\text{[s-d]}}} \big)\odot \big({\bf \Lambda}{\bm \vartheta} {\bm \vartheta}^H + {\bm \vartheta} ({\bf \Lambda}{\bm \vartheta})^H \big)$, and
\begin{equation}\label{III-38-1}
\begin{split}
\frac{\partial \underline{{\bf \Sigma}}^{\text{[r]}}}{\partial {\theta}^{\text{[s-d]}}(m)} =& \big(\alpha^2 \sigma^2_R {\bm \Lambda}_{\bm\theta^{\text{[s-d]}}} \underline{{\bf G}}\underline{{\bf G}}^H  {\bm \Lambda}^H_{\bm\theta^{\text{[s-d]}}} \big) \\
& \odot\big( {\bf a}_{m} [{\bm \theta}^{\text{[s-d]}}]^H +  {\bm \theta}^{\text{[s-d]}}{\bf a}^H_{m}  \big),~\forall m
\end{split}
\end{equation}
where $\boldsymbol\vartheta\triangleq\big[1, e^{\frac{j2\pi \phi^{\text{[s-d]}}}{N}}, \cdots, e^{\frac{j2\pi (N-1)\phi^{\text{[s-d]}}}{N}} \big]$. For channel response $\underline{{\bf g}}$, based on the structure of ${\bf G}$ as shown in \eqref{EQU-10-1}, we have
\begin{equation}\label{III-39}
\begin{split}
\frac{\partial \underline{{\bf \Sigma}}^{\text{[r]}}}{\partial {g(0)}} =& \alpha^2 \sigma^2_R {\bm \Lambda}_{\bm\theta^{\text{[s-d]}}} {\bm \Lambda}_{\phi^{\text{[s-d]}}} \big({\bf D}_0 \underline{{\bf G}}^H + \underline{{\bf G}} {\bf D}^H_0 \big)  {\bm \Lambda}^H_{\phi^{\text{[s-d]}}} {\bm \Lambda}^H_{\bm\theta^{\text{[s-d]}}}, \\
\frac{\partial \underline{{\bf \Sigma}}^{\text{[r]}}}{\partial \Re({g(m)})} =& \alpha^2 \sigma^2_R {\bm \Lambda}_{\bm\theta^{\text{[s-d]}}} {\bm \Lambda}_{\phi^{\text{[s-d]}}} \big( {\bf D}_m \underline{{\bf G}}^H + \underline{{\bf G}} {\bf D}^H_m \big)  \\
& \times {\bm \Lambda}^H_{\phi^{\text{[s-d]}}} {\bm \Lambda}^H_{\bm\theta^{\text{[s-d]}}},~
 m=1,2,\cdots,L_g-1,\\
\frac{\partial \underline{{\bf \Sigma}}^{\text{[r]}}}{\partial \Im({g(m)})} =& j \alpha^2 \sigma^2_R {\bm \Lambda}_{\bm\theta^{\text{[s-d]}}} {\bm \Lambda}_{\phi^{\text{[s-d]}}} \big( {\bf D}_m \underline{{\bf G}}^H -\underline{{\bf G}} {\bf D}^H_m \big) \\
 & \times {\bm \Lambda}^H_{\phi^{\text{[s-d]}}} {\bm \Lambda}^H_{\bm\theta^{\text{[s-d]}}},~
 m=1,2,\cdots,L_g-1,
\end{split}
\end{equation}
where ${\bf D}_m\triangleq\left[{\bf 0}_{N \times (L_g-m-1)}, {\bf I}_N, {\bf 0}_{N \times m}\right]$. Subsequently, the derivatives of the covariance matrix with respect to the relay and imaginary parts of the relay-to-destination channel parameters are given by $\frac{\partial \underline{{\bf \Sigma}}}{\partial {g(0)}} = {\rm Blkdiag}\big(\frac{\partial \underline{{\bf \Sigma}}^{\text{[r]}}}{\partial {g(0)}}, {\bf 0}_{N \times N}\big)$ and $\frac{\partial \underline{{\bf \Sigma}}}{\partial \Im({g(m)})} = {\rm Blkdiag} \big(\frac{\partial \underline{{\bf \Sigma}}^{\text{[r]}}}{\partial \Im({g(m)})}, {\bf 0}_{N \times N}\big)$, respectively. By combing \eqref{III-36-1}-\eqref{III-39} together, the results in Theorem 1 are derived.

\ifCLASSOPTIONpeerreview
\renewcommand{\baselinestretch}{1.5}
\fi

\bibliographystyle{IEEEtran}
\bibliography{IEEEabrv,OFDM_Relay}

\end{document}